\documentclass[useAMS,usenatbib]{mn2e}
\usepackage{graphicx,rotate,url,mathptmx,lscape,times, color,subfigure}
\def\gtsim{\mathrel{\hbox{\rlap{\hbox{\lower4pt\hbox{$\sim$}}}\hbox{$>$}}}}
\def\lesssim{\mathrel{\hbox{\rlap{\hbox{\lower4pt\hbox{$\sim$}}}\hbox{$<$}}}}
\def\Msunpyr{M$_{\odot}\,$yr$^{-1}$}
\def\Msun{M$_{\odot}$}
\def\cm{{\rm\thinspace cm}}
\def\as{{\rm\thinspace arcsec}}
\def\erg{{\rm\thinspace erg}}

\def\Hz{{\rm\thinspace Hz}}

\def\kpc{{\rm\thinspace kpc}}

\def\Msun{\hbox{$\rm\thinspace M_{\odot}$}}

\def\s{{\rm\thinspace s}}

\def\W{{\rm\thinspace W}}
\def\yr{{\rm\thinspace yr}}

\def\ergpscmpspsas{\hbox{$\erg\cm^{-2}\s^{-1}\as^{-2}\,$}}

\def\Msunpyr{\hbox{$\Msun\yr^{-1}\,$}}

\def\WpHz{\hbox{$\W\Hz^{-1}\,$}}
\def\h0{\hbox{{\rm H}$^0$}}
\DeclareMathAlphabet{\vib}{OML}{cmm}{m}{it}

\def\MRC{MRC\,0406-244}

\def\rf{$R_{\rm 606}$}
\def\iff{$I_{\rm 814}$} 
\def\j{$J_{\rm 110}$}
\def\h{$H_{\rm 160}$}

\def\GALFIT{{GALFIT}}

\begin{document}
\title[The host galaxy of \MRC]{The host galaxy of the $z=2.4$ radio-loud AGN \MRC\ as seen by {\it HST}}

\author[N.\,A.\,Hatch et al.]
       {\parbox[]{6.0in}
       {N.\,A.\,Hatch$^{1}$\thanks{E-mail: nina.hatch@nottingham.ac.uk},  H.\,J.\,A.\,R\"ottgering$^2$, G.\,K.\,Miley$^2$, E.\,Rigby$^2$, C.\,De\,Breuck$^3$, H.\,Ford$^4$, E.\,Kuiper$^2$, J.\,D.\,Kurk$^5$, R.\,A.\,Overzier$^{6,7}$, L.\,Pentericci$^8$.\\        \footnotesize
        $^1$School of Physics and Astronomy, University of Nottingham, University Park, Nottingham NG7 2RD\\
        $^2$Leiden Observatory, Leiden University, P.B. 9513, Leiden 2300 RA, The Netherlands\\
	$^3$European Southern Observatory, Karl-Schwarzschild-Str. 2, D-85748 Garching, Germany\\
	$^4$Department of Physics and Astronomy, Johns Hopkins University, Baltimore, MD, USA\\ 
	$^5$Max-Planck-Institut fuer Extraterrestrische Physik, Giessenbachstrasse, D-85748 Garching, Germany\\  
	$^6$Department of Astronomy, University of Texas at Austin, C1400, 1, University Station, Austin, TX 78712, USA\\ 
	$^7$Observat\'orio Nacional, Rua Jos\'e Cristino, 77. CEP 20921-400, S\~ao Crist\'ov\~ao, Rio de Janeiro-RJ, Brazil\\  
	$^8$INAF, Osservatorio Astronomica di Roma, Via Frascati 33, 00040 Monteporzio, Italy\\      
    }}
 \date{}
\pubyear{}
\maketitle

\label{firstpage}
\begin{abstract}
We present multicolour Hubble Space Telescope images of the powerful $z=2.4$ radio galaxy \MRC\ and model its complex morphology with several components including a host galaxy, a point source, and extended nebular and continuum emission. We suggest that the main progenitor of this radio galaxy was a normal, albeit massive (M$_\star\sim10^{11}$\Msun), star-forming galaxy. The optical stellar disc of the host galaxy is smooth and well described by a S\'ersic profile, which argues against a recent major merger, however there is also a point-source component which may be the remnant of a minor merger. The half-light radius of the optical disc is constrained to lie in the range $3.5$ to $8.2$\,kpc, which is of similar size to coeval star forming galaxies. 

Biconical shells  of nebular emission and UV-bright continuum extend out from the host galaxy along the radio jet axis, which is also the minor axis of the host galaxy. The origin of the continuum emission is uncertain, but it is most likely to be young stars or dust-scattered light from the AGN, and it is possible that stars are forming from this material at a rate of $200^{+1420}_{-110}$\Msunpyr.

\end{abstract}
\section{Introduction}

\begin{figure}
\centering
\includegraphics[angle=90,height=1\columnwidth]{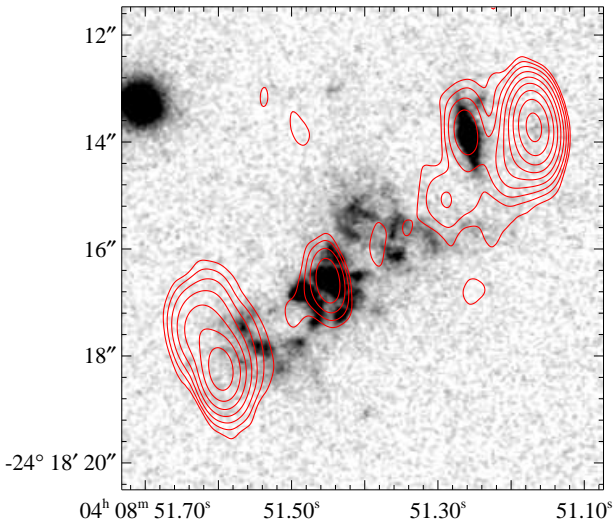}
\caption{A {\it HST} \rf\ image of \MRC\ with 4.71\,GHz radio emission from  \citet{Carilli1997} overlaid in red contours. Contours mark the radio emission at 0.17$\times$ [1, 2, 4, 8, 16, 32, 64, 128, 256] mJy\,beam$^{-1}$. The object which overlaps the northwest radio lobe is a foreground galaxy that is not part of \MRC\ \citep{Rush1997}. \label{MRC0406_radio}}
\end{figure}

Distant powerful radio galaxies ($z > 2$, $L_{500\,{\rm MHz~(rest)}}>10^{27}$\WpHz) pinpoint rapidly accreting supermassive black holes. The observational evidence suggests these black holes are situated in massive galaxies \citep{Rocca-Volmerange2004, Seymour2007}, some of which are growing at rates of hundreds to thousands of solar masses per year \citep{Seymour2012,Barthel2012}. Such galaxies are particularly interesting because they are undergoing active galactic nuclei (AGN) feedback in which a significant fraction of the ambient gas in their haloes could become unbound \citep{Nesvadba2006,Nesvadba2008}. Cosmological simulations suggest that these feedback episodes in the early Universe may play an important role in galaxy evolution \citep{diMatteo2005,Booth2009}, so it is interesting to determine what kind of galaxy they occur in, what triggers these events and how they affect the host galaxy.

Most host galaxies of  $1<z<3$  obscured AGN are disc-dominated and few contain signs of recent major mergers, which suggests that secular processes drive black hole growth for most of the time \citep{Schawinski2012,Simmons2012}. However, the most luminous phases of AGN activity seem to be triggered by major mergers \citep{Treister2012}.  Radio galaxies at $z>2$ appear clumpy and extended  \citep[e.g.][]{Pentericci1999,Pentericci2001}, and sub-millimetre observations suggest that gas-rich mergers may play the dominant role in fuelling their central black holes \citep{Ivison2012}. 

We selected the radio galaxy  \MRC\ at $z=2.44$ from the original sample of \citet{Pentericci2001} for a more detailed study. Its particularly complex morphology may reveal what triggered the AGN or be indicative of feedback processes. This galaxy consists of multiple clumpy components, and is undergoing powerful, jet-induced feedback \citep{Nesvadba2008}. A high resolution 4.71\,GHz radio map of \MRC\ is displayed Fig.\,\ref{MRC0406_radio}: a central radio core is flanked by two hot-spots at a position angle of $128^\circ$, which are separated by 61\,kpc (7.4\arcsec) and lie beyond the rest-frame optical and ultraviolet (UV) emission. The optical and UV emission, as well as the large Ly$\alpha$ halo are aligned with the  radio hot-spots, but there is no spatial overlap.  

\MRC\ is a well studied galaxy \citep[e.g.][]{Rush1997,Pentericci2001,Taniguchi2001,Humphrey2009}. Hubble Space Telescope ({\it HST}) images revealed several connecting bright clumps in a figure-of-eight shape elongated along the radio jet axis. \citet{Rush1997} concluded that the complex morphology was of tidal origin and indicative of a recent merger event. On the other hand,  \citet{Taniguchi2001}  and \citet{Humphrey2009} argued that the morphology resembled the bubbles found in ultra-luminous  infrared galaxies, favouring a formation mechanism for \MRC\ that involves a superwind from a starburst event. These works also showed that the extended structure was dominated by nebular emission (which was excited by a hidden AGN), although UV continuum emission was also observed in the same regions and was presumed to follow the same morphology. \citet{Humphrey2009} argue that the high velocities of the emission line gas suggest that they are transient phenomena that will escape the gravitational pull of the galaxy and disperse into the intergalactic medium on the timescale of 100\,Myrs.

We expand on these earlier studies by performing a multi-wavelength morphological study. Using high-resolution optical and near-infrared {\it HST} images, we separate the light from \MRC\ into its constituent parts: the host galaxy, a point-source, nebular emission, and extended continuum.  We determine the morphology of the host galaxy and derive its mass and star formation rate through fitting its spectral energy distribution (SED). We compare these properties to similarly massive galaxies in the literature to determine what type of galaxy \MRC\ was before the AGN activity, and what effect AGN feedback has on the host galaxy. 

In Section \ref{observations} we describe our observations and how we create the nebular and continuum images from the broad-band {\it HST} data. In Section \ref{analysis} we use \GALFIT\ to model the morphology of \MRC\ using several components, and analyse their SEDs. Finally, we discuss the current state and speculate on the future evolution of \MRC\ in Section \ref{discussion}. We use AB magnitudes throughout, and assume a \citet{Calzetti2000} extinction law and a $\Lambda$CDM cosmology with H$_0$ = 71, $\Omega_m = 0.27$ and $\Omega_\Lambda= 0.73$. 

\section{Nebular and continuum images} 
\label{observations}
\subsection{Observations}

\begin{figure*}
\begin{minipage}[h]{52mm}
\centering
\resizebox{52mm}{!}{
\includegraphics[angle=90,height=1\columnwidth]{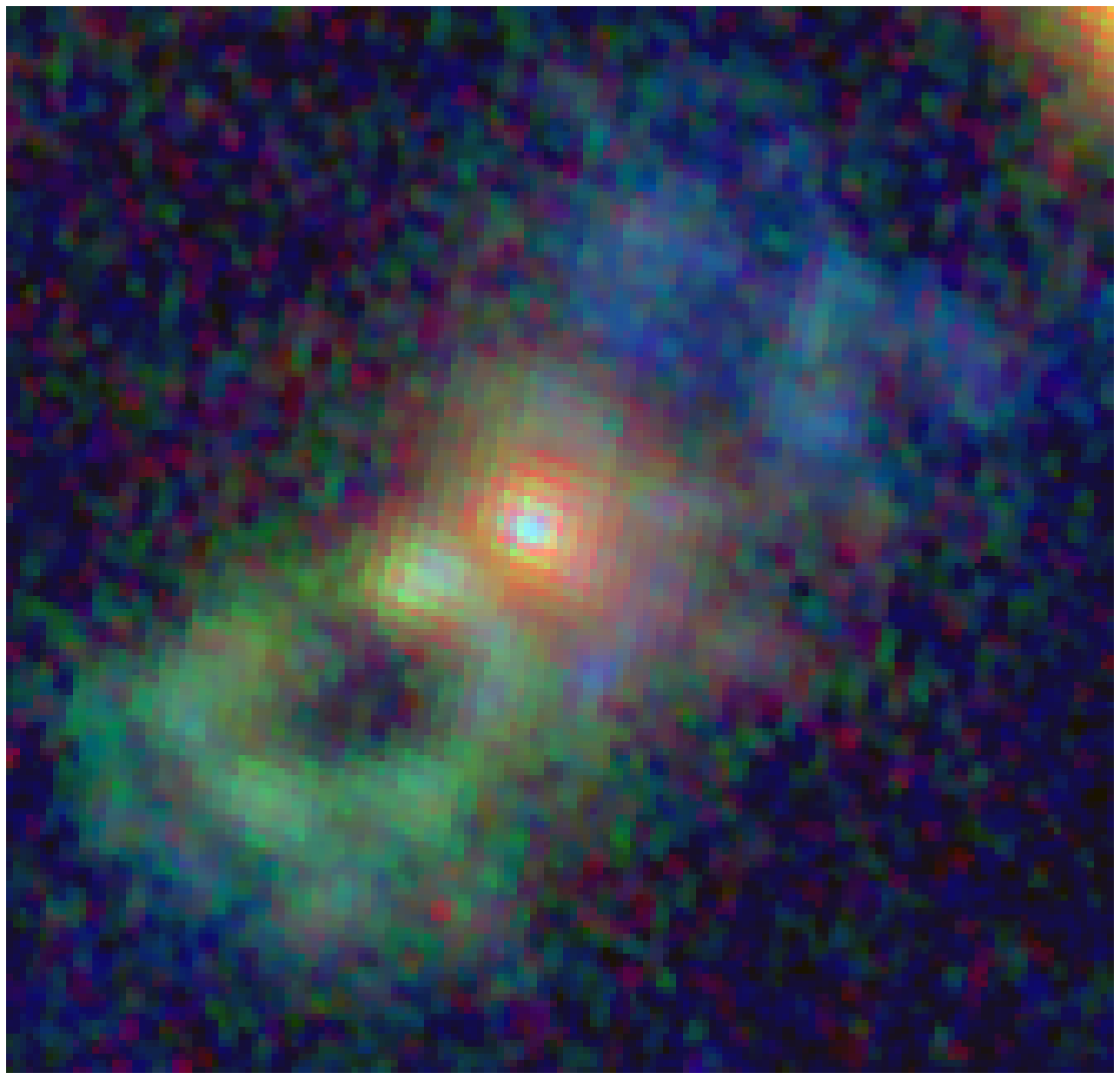}}
\end{minipage}
\begin{minipage}[h]{124mm}
\centering
\resizebox{124mm}{!}{
\includegraphics[height=2.5cm]{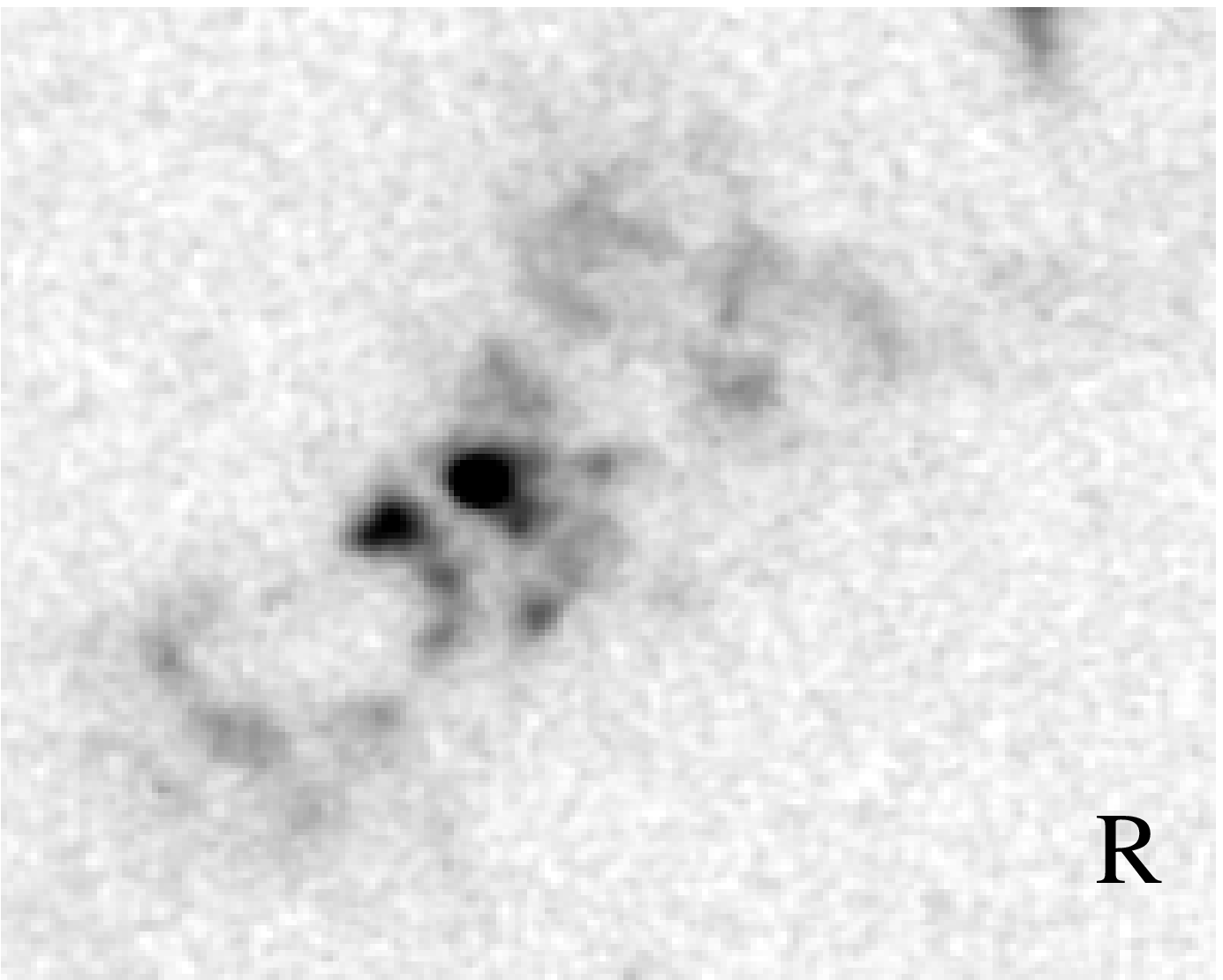}
\includegraphics[height=2.5cm]{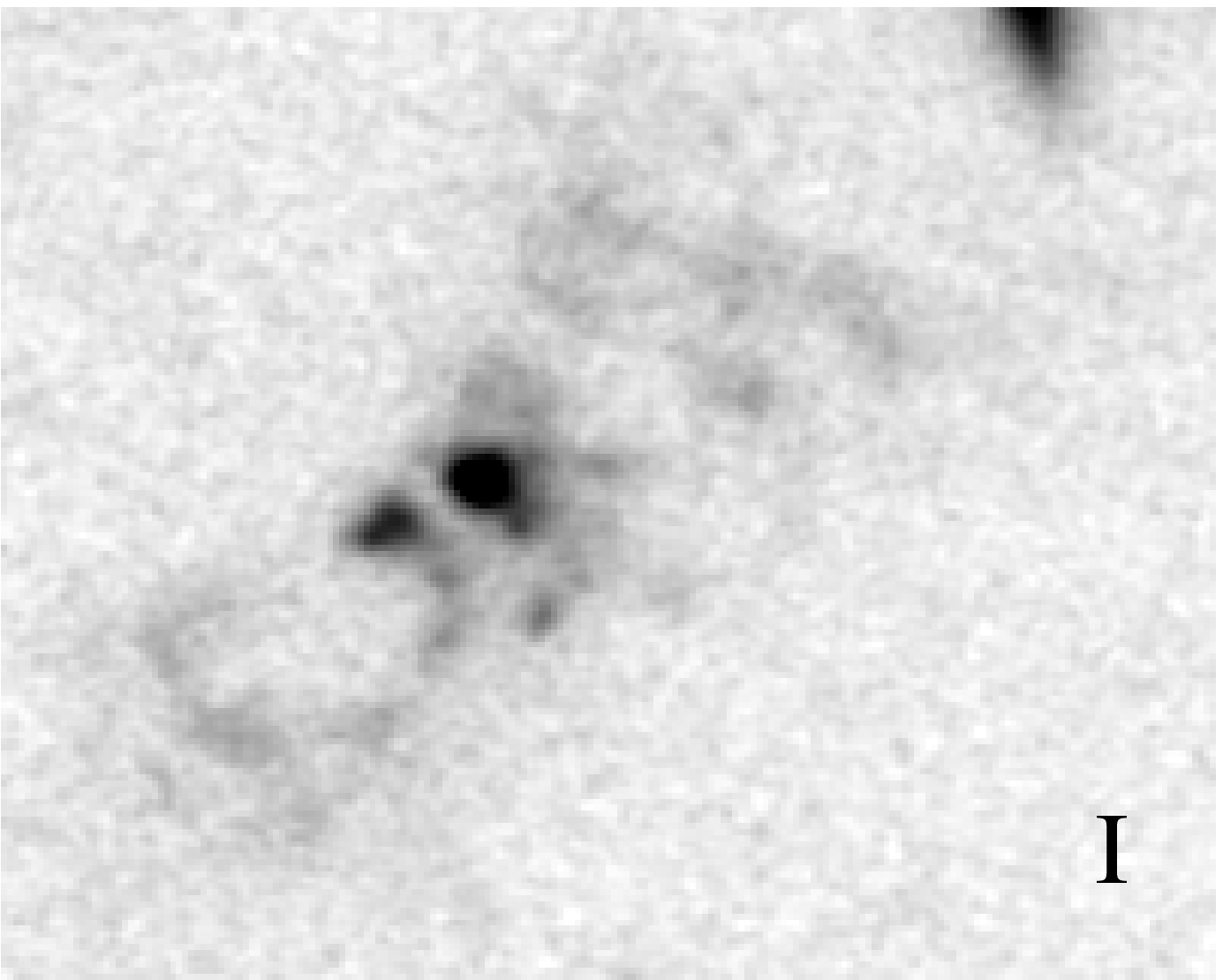}
\includegraphics[height=2.5cm]{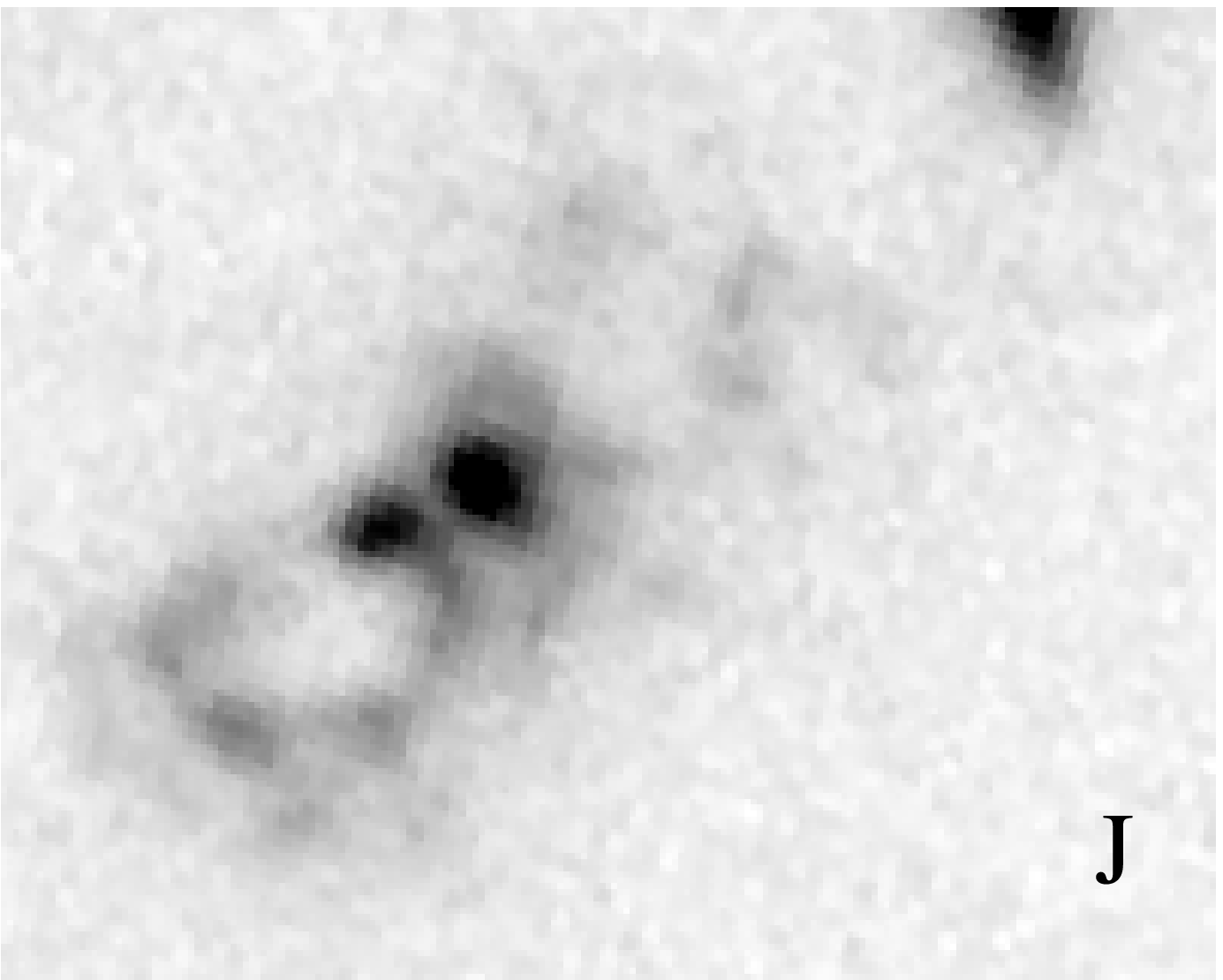}
\includegraphics[height=2.5cm]{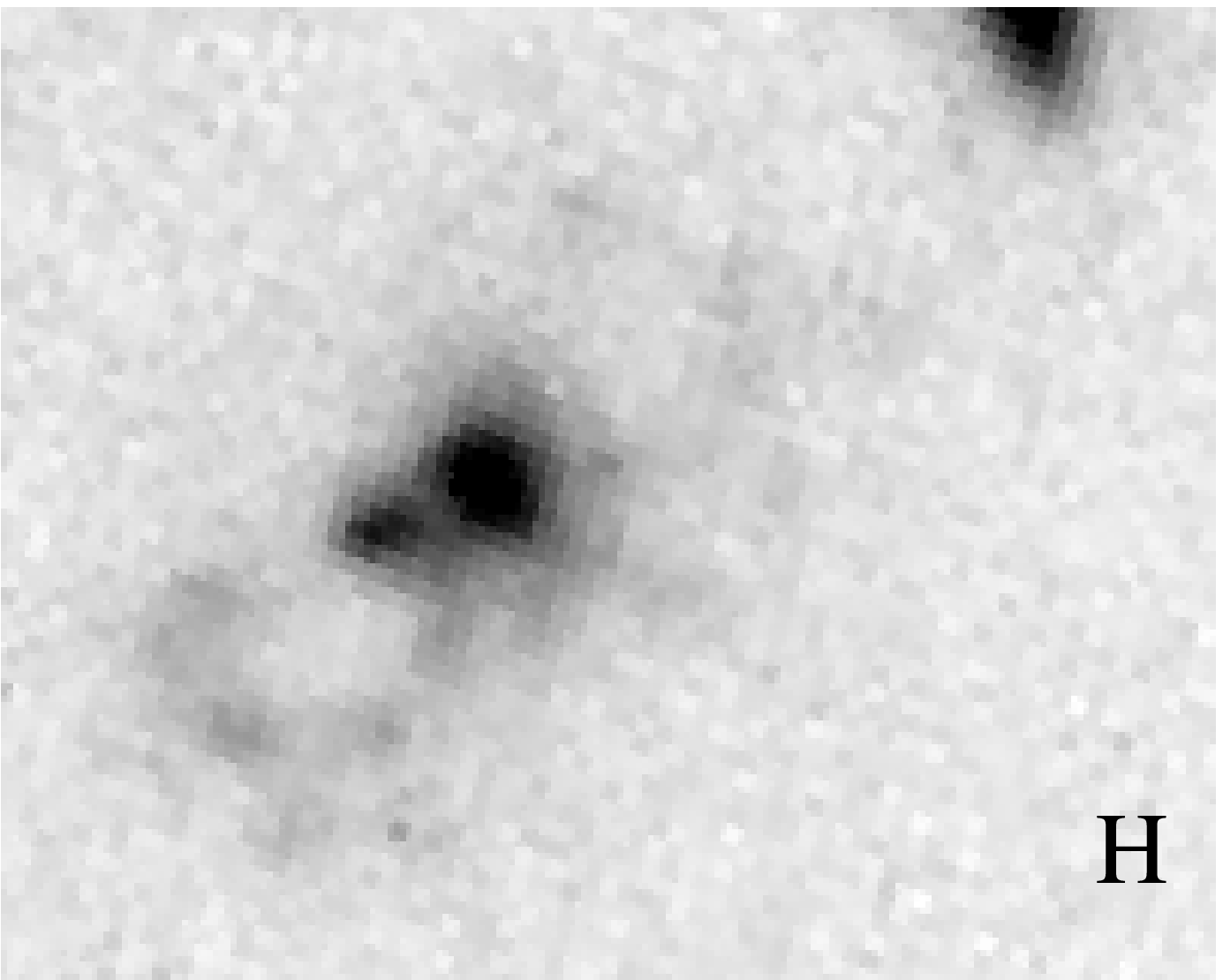}
}
\resizebox{124mm}{!}{
\includegraphics[height=2.5cm]{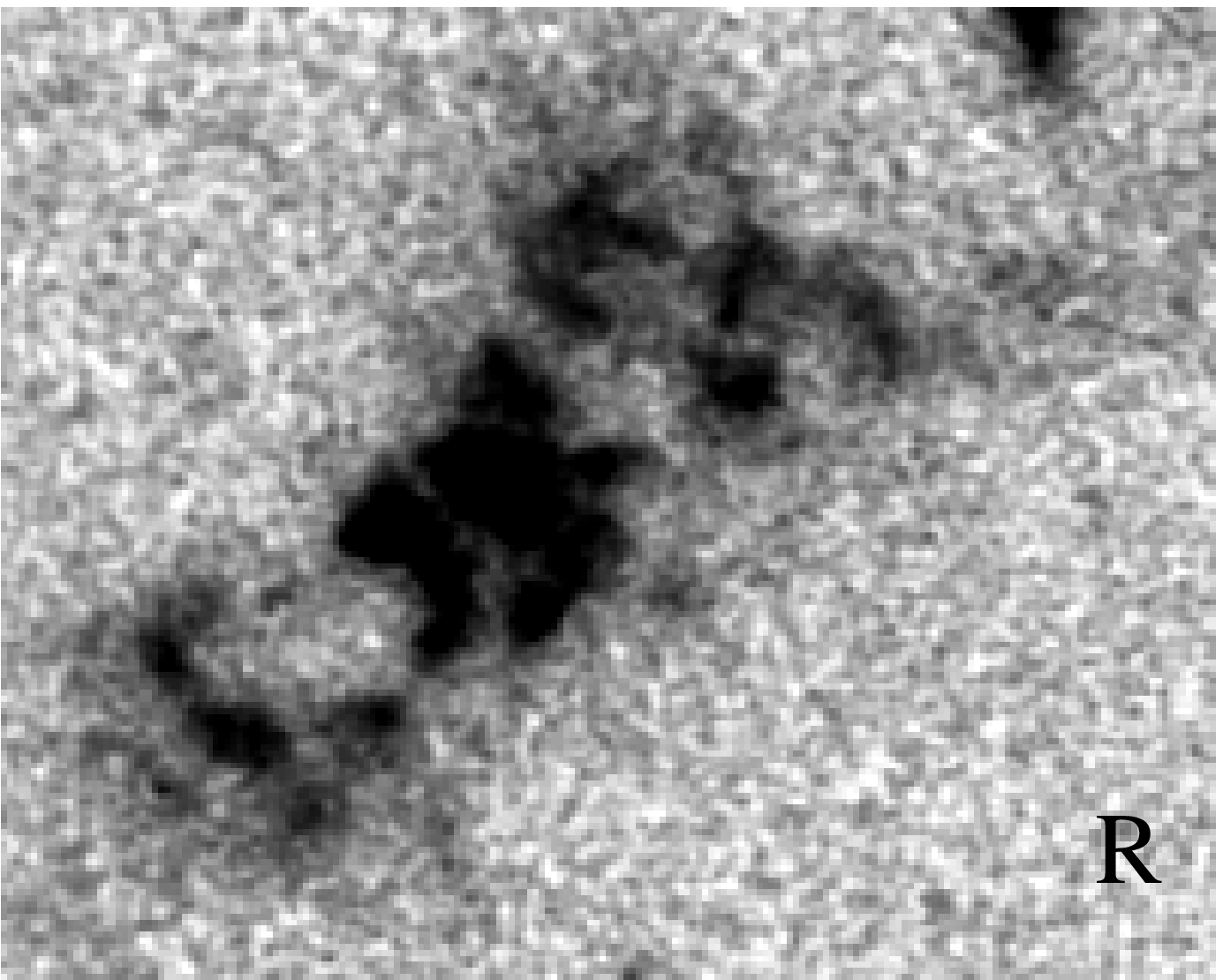}
\includegraphics[height=2.5cm]{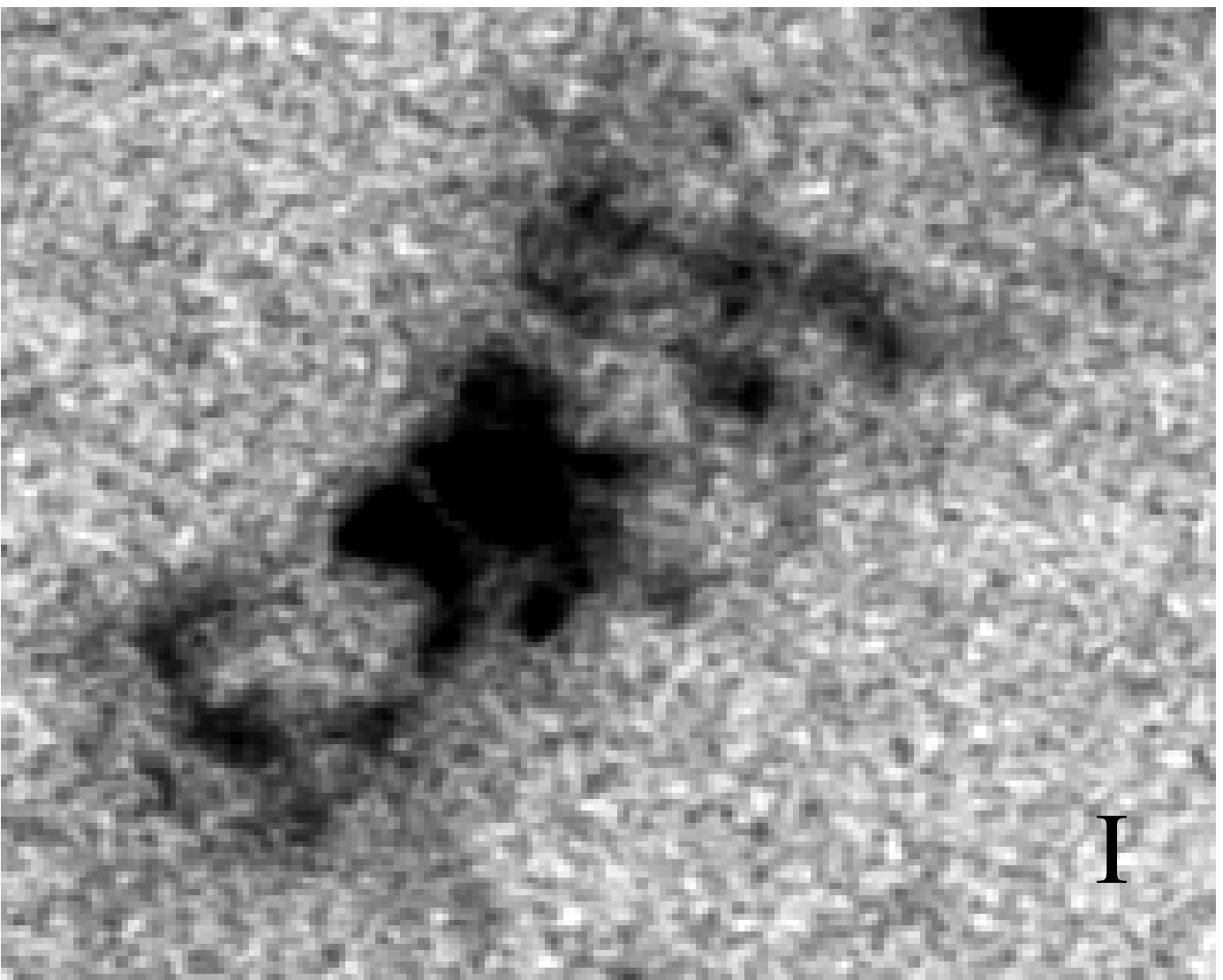}
\includegraphics[height=2.5cm]{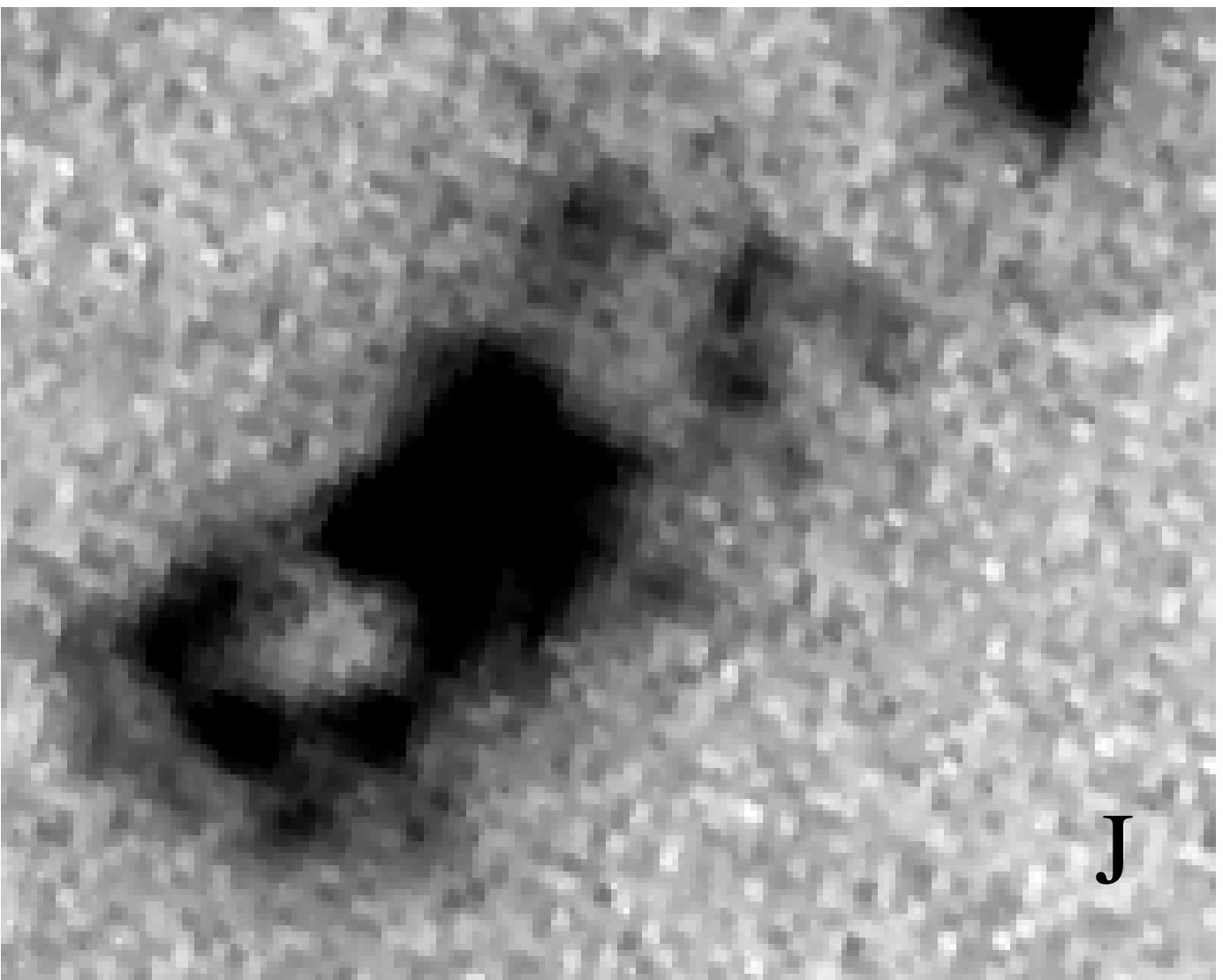}
\includegraphics[height=2.5cm]{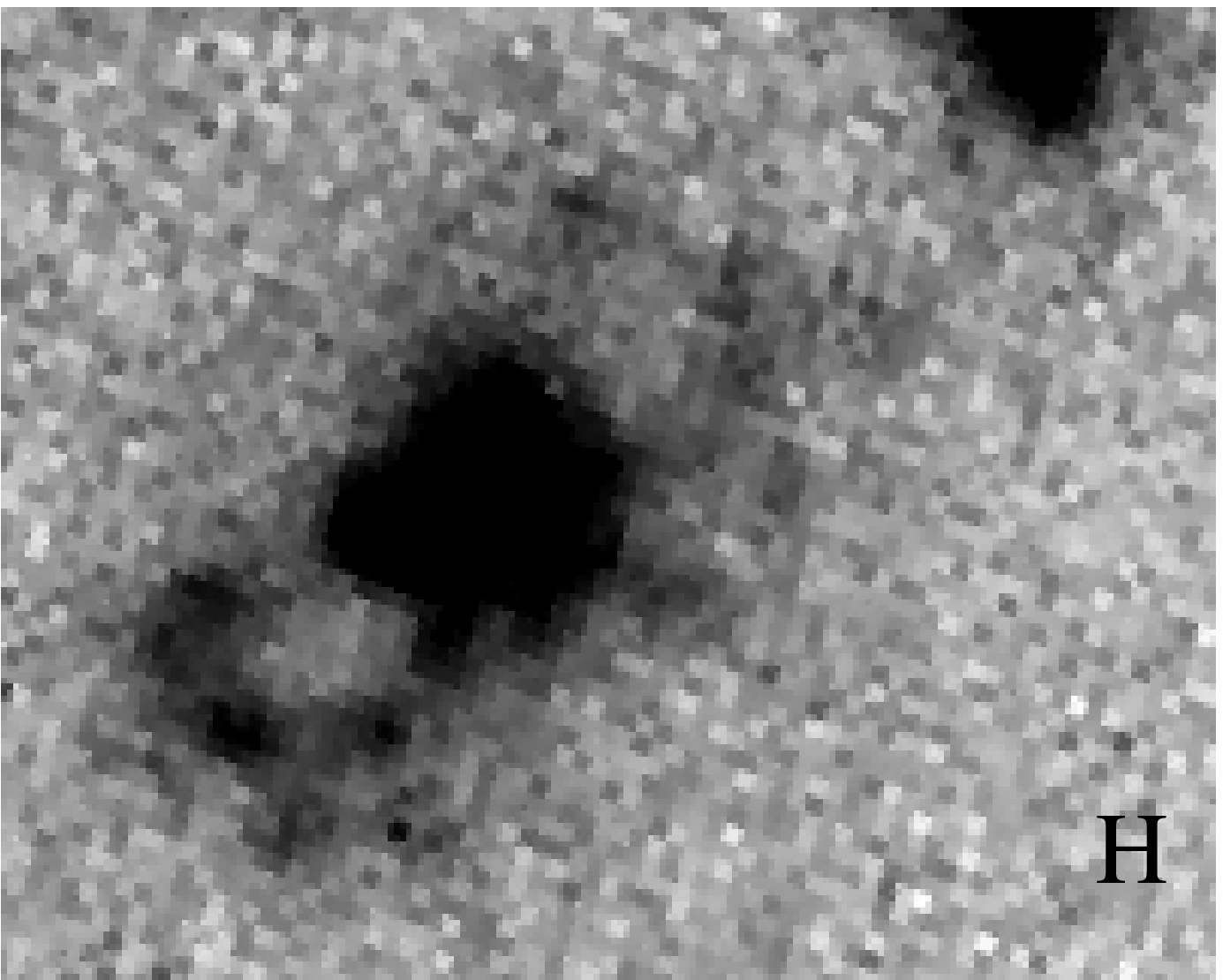}
}
\end{minipage}

\caption{Colour image of \MRC\ (left) with  blue -- \rf, green -- \j\ and red -- \h, and full resolution \rf, \iff, \j\ and \h\ images (right) at two different contrasts to highlight structure at different flux levels. The colour image is $5.0\arcsec \times 4.8$\arcsec\ ($41 \times 40$\,kpc), and North is up and East is left in all images.  The object at the top right of the \MRC\ image is a foreground galaxy that is not part of \MRC\ \citep{Rush1997}. \label{MRC0406_1}}
\end{figure*}

Images of \MRC\ at $z=2.44$ were obtained under program 11738 (PI.~G.K.\,Miley) with the Advanced Camera for Surveys (ACS) and the Wide Field Camera 3 (WFC3) on board {\it HST}. A single $3.5\arcmin\times3.4$\arcmin\ ACS and $2.3\arcmin\times2.1$\arcmin\ WFC3 field was observed through the optical filters F606W (\rf ) and F814W (\iff ) with ACS, and near-infrared filters F110W (\j ) and F160W (\h) with WFC3. The total exposure time was 1 orbit in the \j\ and \h\ filters (2612 seconds), and 4 orbits in each of the \rf\ and \iff\ filters (10173 seconds each). The observations were reduced using the {\sc pyraf} task {\sc multidrizzle} to produce cosmic-ray removed, registered images with a common scale of 0.04\arcsec per pixel.  Images were corrected for Galactic extinction using an $R_{V}=3.1$ extinction law and $E{\rm(}B-V{\rm )}=0.049$\,mag  \citep{SchlaflyFinkbeiner2011}. The reduced images and a false three-colour [\rf, \j, \h] image of the radio galaxy are displayed in Fig.\,\ref{MRC0406_1}. \MRC\ comprises several large clumps and filaments extending approximately 36\,kpc (4.4\arcsec) along the same orientation as the radio jets. There is a strong asymmetry in the colour of the northern and southern extensions. 

\subsection{Nebular emission within filter passbands}
\MRC\ is a strong line emitter and all of the broadband images encompass some nebular emission. We estimated the amount of light contributed by nebular emission within each observed filter using the long-slit spectroscopy of \citet{Taniguchi2001}, \citet{Iwamuro2003} and \citet{Humphrey2008}.
Both the continuum flux density and line fluxes were taken from the same data set where possible. This was not possible for the \rf\ and \iff\ bands so instead the continuum was measured directly from the {\it HST} images from an area matched to the slit size.

The nebular continuum contribution was calculated with the {\sc dipso} {\sc starlink} code {\sc nebcont} using the H$\beta$ flux. We assume both temperature (11,700$^{+1800}_{-1500}$\,K) and electron density ($500^{+500}_{-200}$\,cm$^{-3}$) are constant across the nebula \citep{Nesvadba2008,Humphrey2008}. The H$\beta$ flux was first corrected for extinction ($A_V=0.3$\,mag; \citealt{Humphrey2008}), and then the nebular continuum reddened using the same extinction values to allow comparison to the observed continuum. 

In Table \ref{nebular} we list the estimated nebular contribution to each filter. These estimates are used in Section~\ref{neb_sub} in combination with the [O{\sc iii}] image (derived below) to produce continuum images from the {\it HST} data.

\subsection{[O{\sc iii}] emission image} 
\begin{figure}

\includegraphics[width=1\columnwidth]{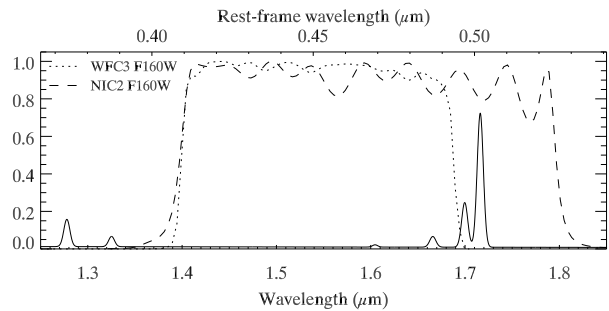}

\caption{Filter transmission curves of F160W filters on NICMOS 2 (dashed line) and WFC3 (dotted line) instruments overlaid on a near-infrared spectrum of \MRC\ (line and continuum values taken from \citealt{Iwamuro2003}). The NICMOS-F160W filter covers the bright [O{\sc iii}] doublet at 4959/5007\AA\ which is avoided by the WFC3 \h\ filter. \label{filter}}

\end{figure}

\begin{figure*}

\begin{minipage}[h]{176mm}
\centering
\resizebox{176mm}{!}{
\includegraphics[height=2.5cm]{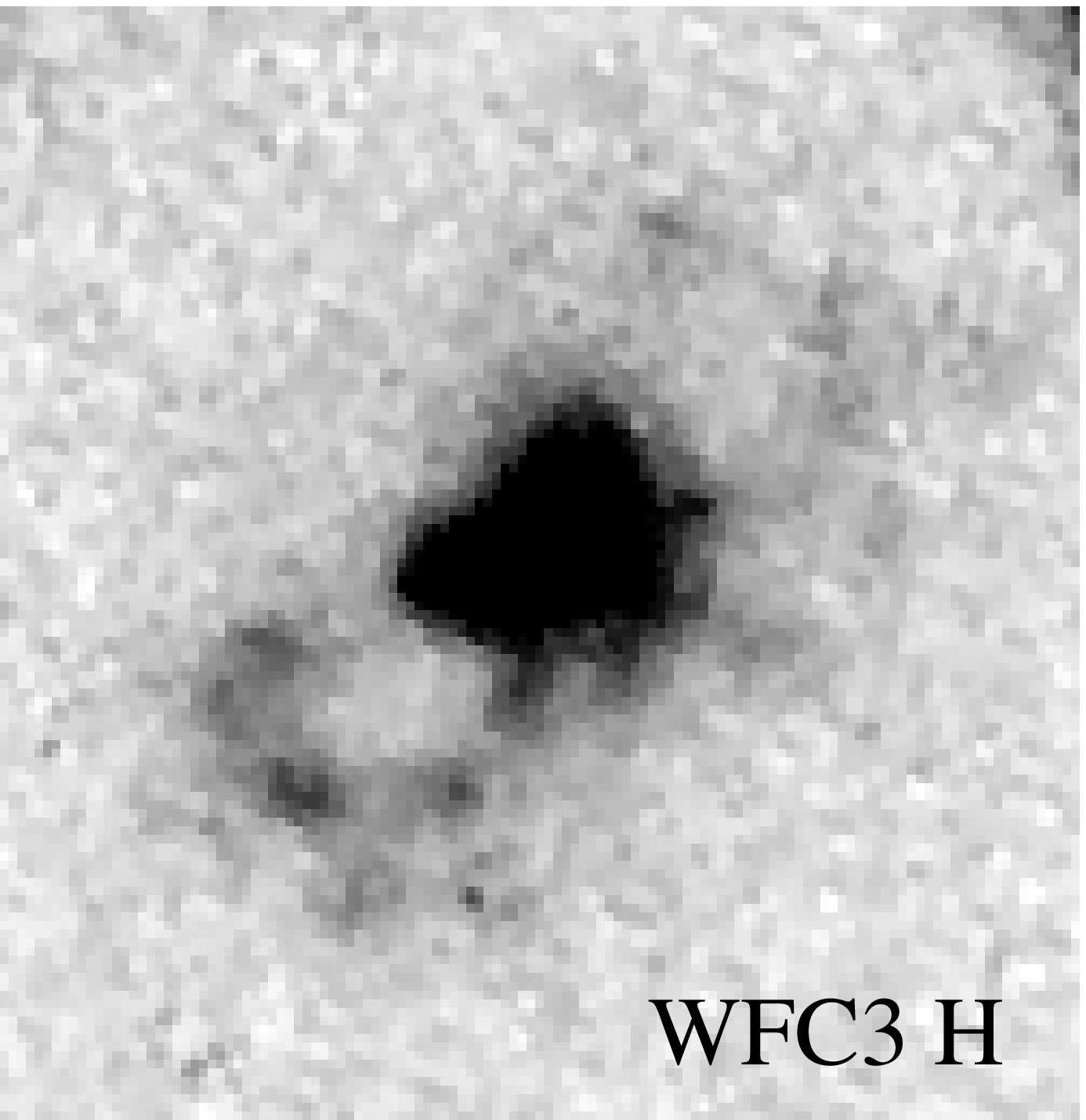}
\includegraphics[height=2.5cm]{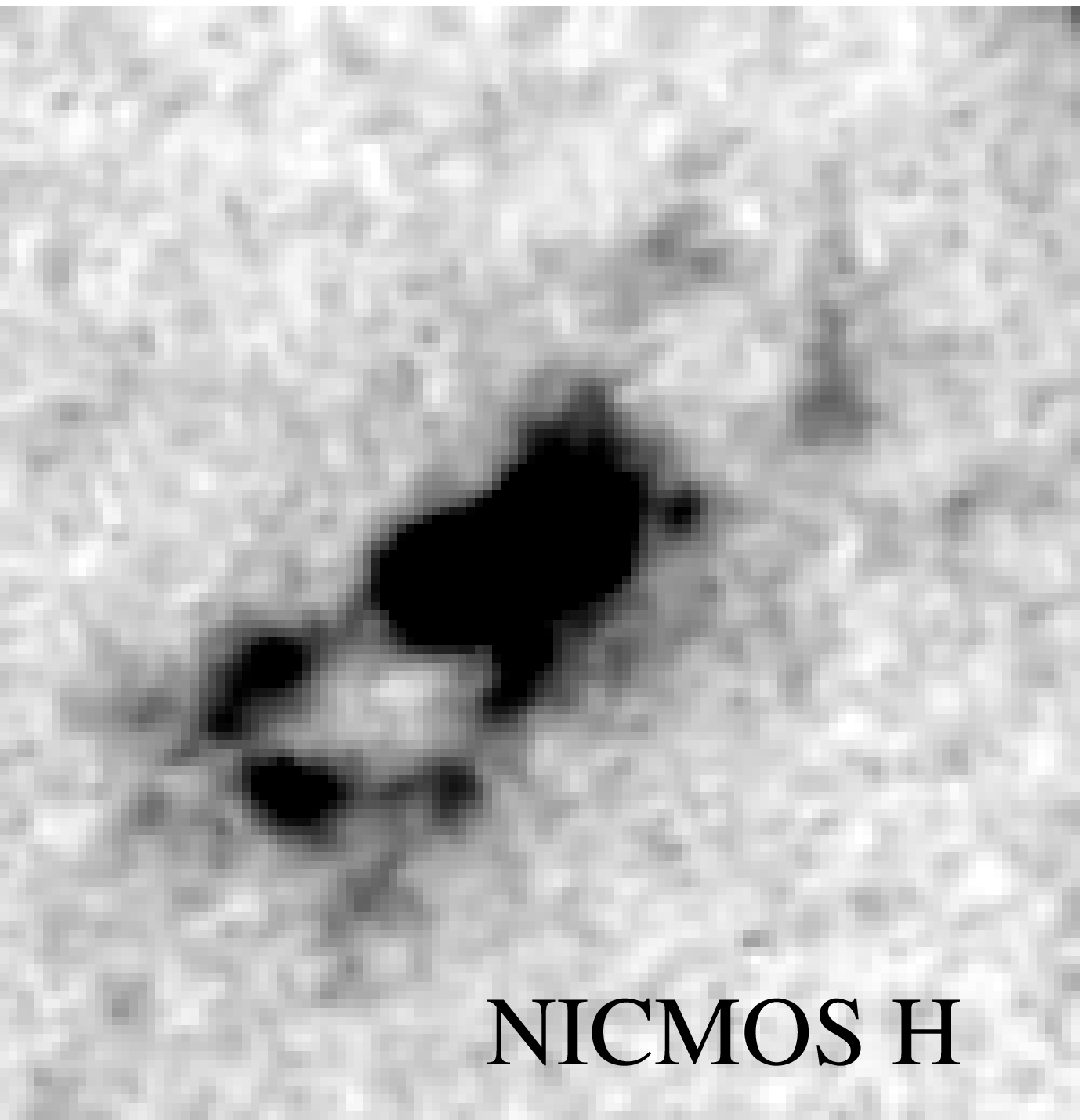}
\includegraphics[height=2.5cm]{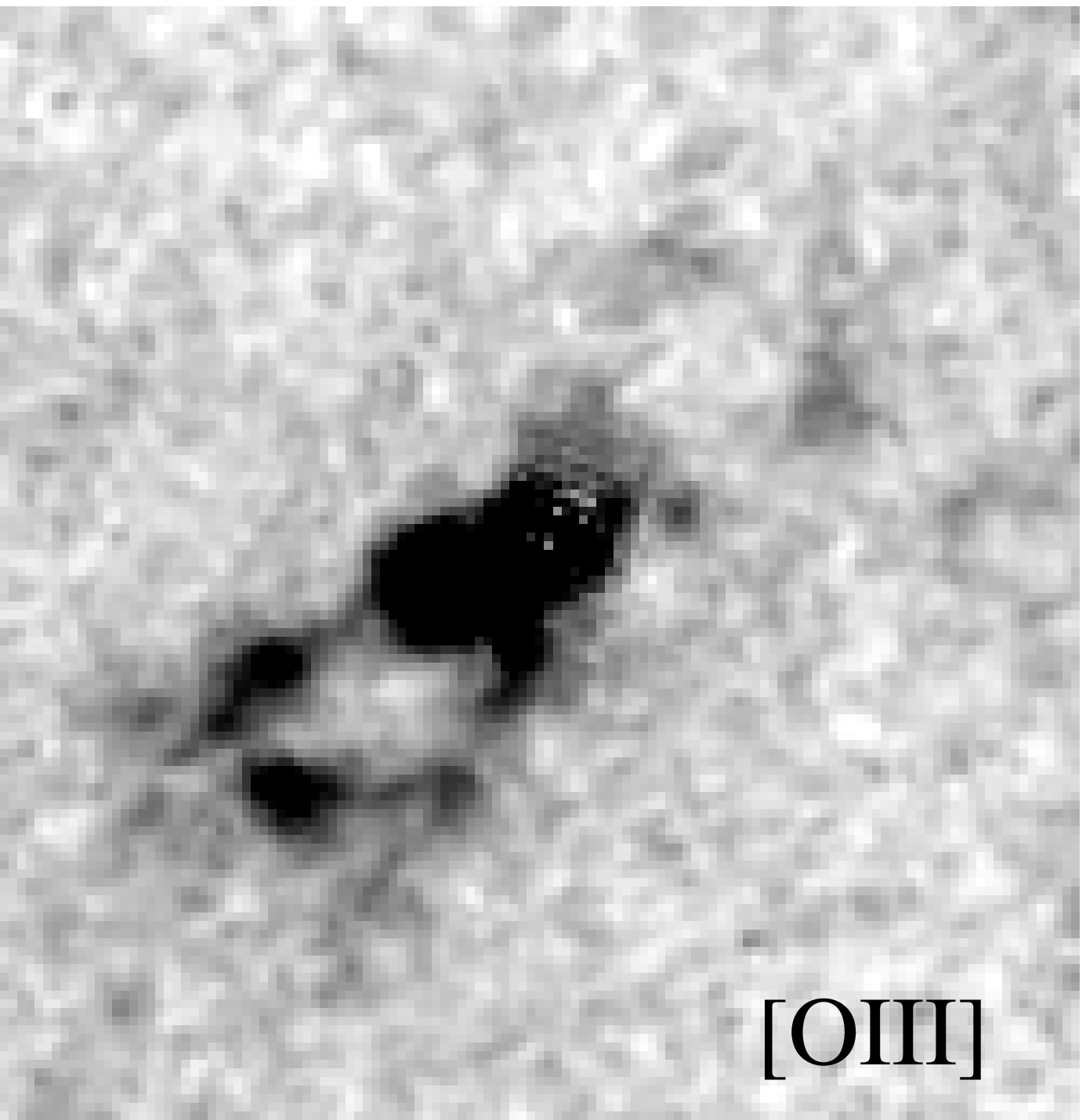}
}
\end{minipage}
\caption{From left to right: the original {\it HST} WFC3 \h\ and  NIC2 \h\ images, and the derived nebular images: [O{\sc iii}] and H$\beta+$nebular continuum. The [O{\sc iii}] image was derived by subtracting the WFC3 \h\ image from the NIC2 \h\ image. All images are $4.9\arcsec\times5.2\arcsec$ (40.5\kpc$\times$42.5\,kpc) and North is up, East is left. \label{MRC_nebular}}

\end{figure*}

\MRC\ was observed through the \h\ filter with WFC3 and previously with NICMOS \citep{Pentericci2001}. The NICMOS-F160W filter (NIC2) is marginally wider than the WFC3 \h\ filter so it  encompassed the bright [O{\sc iii}] doublet at 4959/5007\AA\ that is not covered by the WFC3 \h\ filter (see Fig.\,\ref{filter}). Subtracting the WFC3 emission from the NICMOS image shows the distribution of [O{\sc iii}] emission surrounding \MRC. The resulting map of [O{\sc iii}] emission is shown in Fig.\,\ref{MRC_nebular}. 

Details on the reduction of the NIC2 image can be found in \citet{Pentericci2001}. The NIC2 image was registered and flux calibrated to match the WFC3 image.  The difference in the filter profiles means the galaxy's magnitudes measured in the NIC2 image are slightly brighter than those measured from the WFC3 image (${\rm NIC2}-{\rm WFC3} = -0.1(J-H)-0.03$ for $z<2 $ galaxies), which was taken into account when flux calibrating the NIC2 image. 
We used {\sc Tiny-Tim} to compare the NIC2 and WFC3 F160W PSFs and found them similar so we do not apply a PSF correction to the data. 

For \MRC\ at $z=2.44$  we find ${\rm NIC2}-{\rm WFC3} = -0.01$\,mag by linearly extrapolating the continuum measurements from \citet{Iwamuro2003} and convolving it with the filter profiles. This small difference in continuum level will not produce a significant continuum residual in our derived [O{\sc iii}] image as the [O{\sc iii}]$\lambda4959,5007$ lines have an observed equivalent width of 7673\AA\  \citep{Iwamuro2003} compared to the 947\AA\ difference in filter profiles. 

\begin{table*}
\caption{\label{nebular} Contribution from nebular emission. Values in parentheses give the contribution of nebular continuum and line emission, respectively.  [Ne{\sc iv}]\,2429 lie within the \iff\ band and  [Ne{\sc v}]\,3426 and [Ne{\sc iii}]\,3869 lie within the \j\ band. However, these high-excitation lines have not been included in our estimates as a large fraction of the flux is likely to be emitted from a small region around the central AGN and we are more interested in the extended emission line region. Approximately 3\% of the \j\ flux may come from [NeV]3426 and [NeIII]3869, and 5\% of the \iff\ flux may come from [NeIV]2429, thus these fractional fluxes are within the uncertainties.}
 \centering 
\begin{tabular}{|l|l|l|l|l|l|}
  \hline
band &	\rf\	&	\iff\	&	\j\	&	\h\	\\ \hline
Contrib. (cont., lines)&40$\pm7$\% (21$\pm6$\%, 19$\pm1$\%)		&	30$\pm11$\%	 (22$\pm8$\%, 8$\pm3$\%)&	39$\pm8$\%	 (20$\pm5$\%, 19$\pm3$\%)	&	29$\pm4$\%	 (8$\pm2$\%, 21$\pm2$\%)	\\ 
Emission lines& C{\sc iv}1549, He{\sc ii}1640, C{\sc iii}]1909,  O{\sc iii}]1663 & C{\sc ii}]2326,  [O{\sc ii}]2471 &Mg{\sc ii}2798,  [O{\sc ii}]3727& H$\gamma$, He{\sc ii}4686,  H$\beta$ \\ \hline
\end{tabular}
\end{table*}

\begin{figure*}

\begin{minipage}[h]{176mm}
\centering
\resizebox{176mm}{!}{
\includegraphics[height=2.5cm]{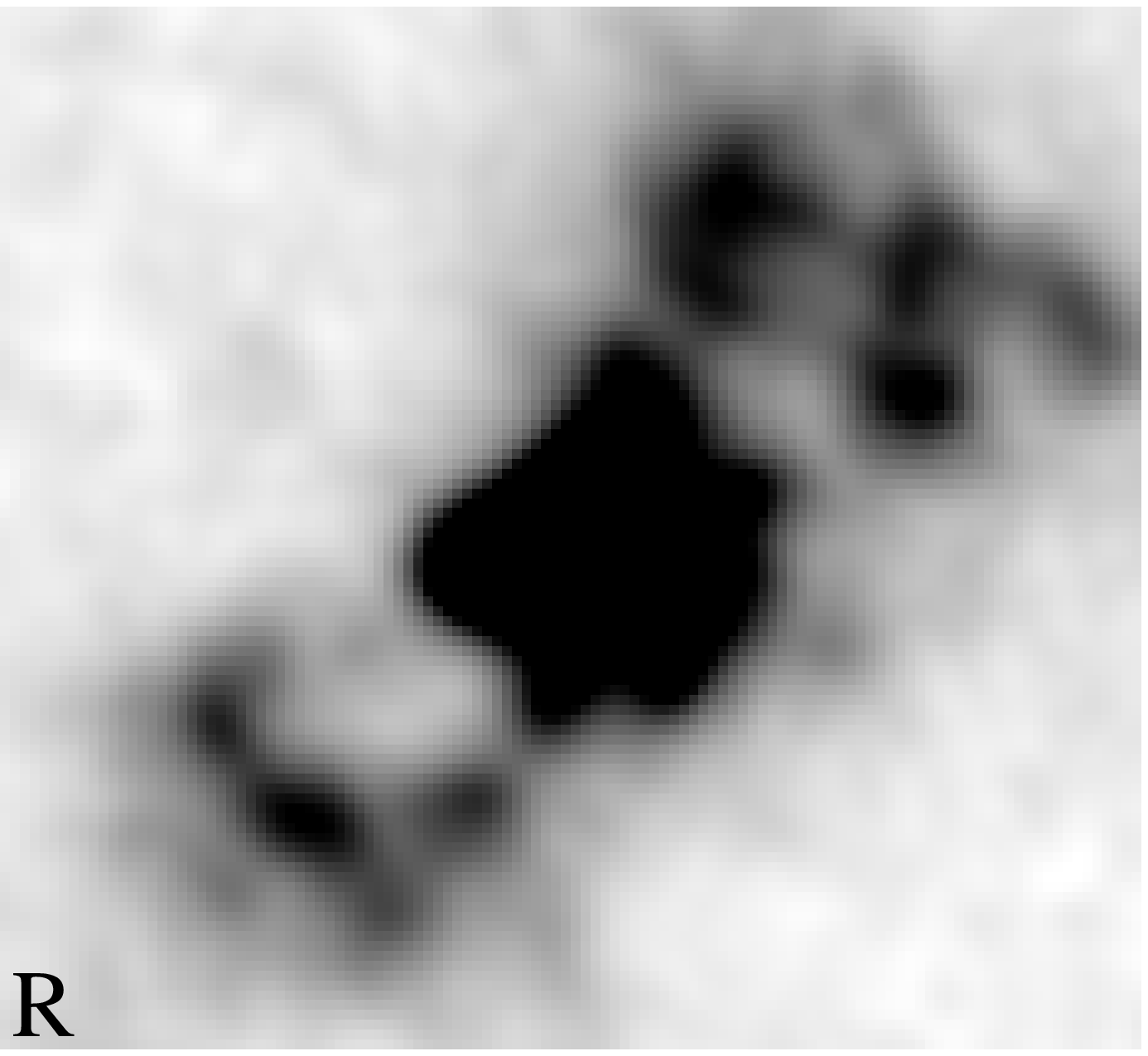}
\includegraphics[height=2.5cm]{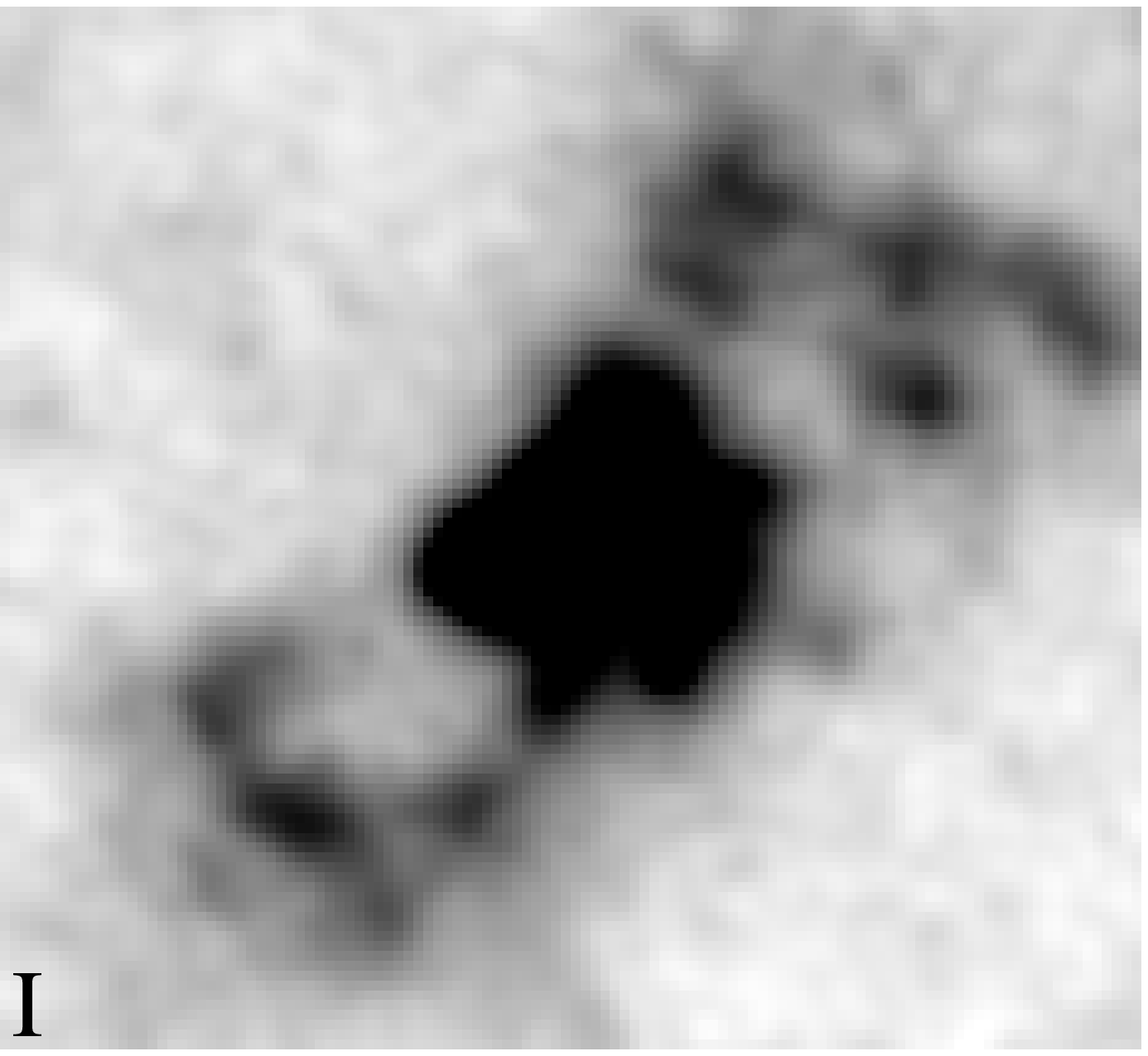}
\includegraphics[height=2.5cm]{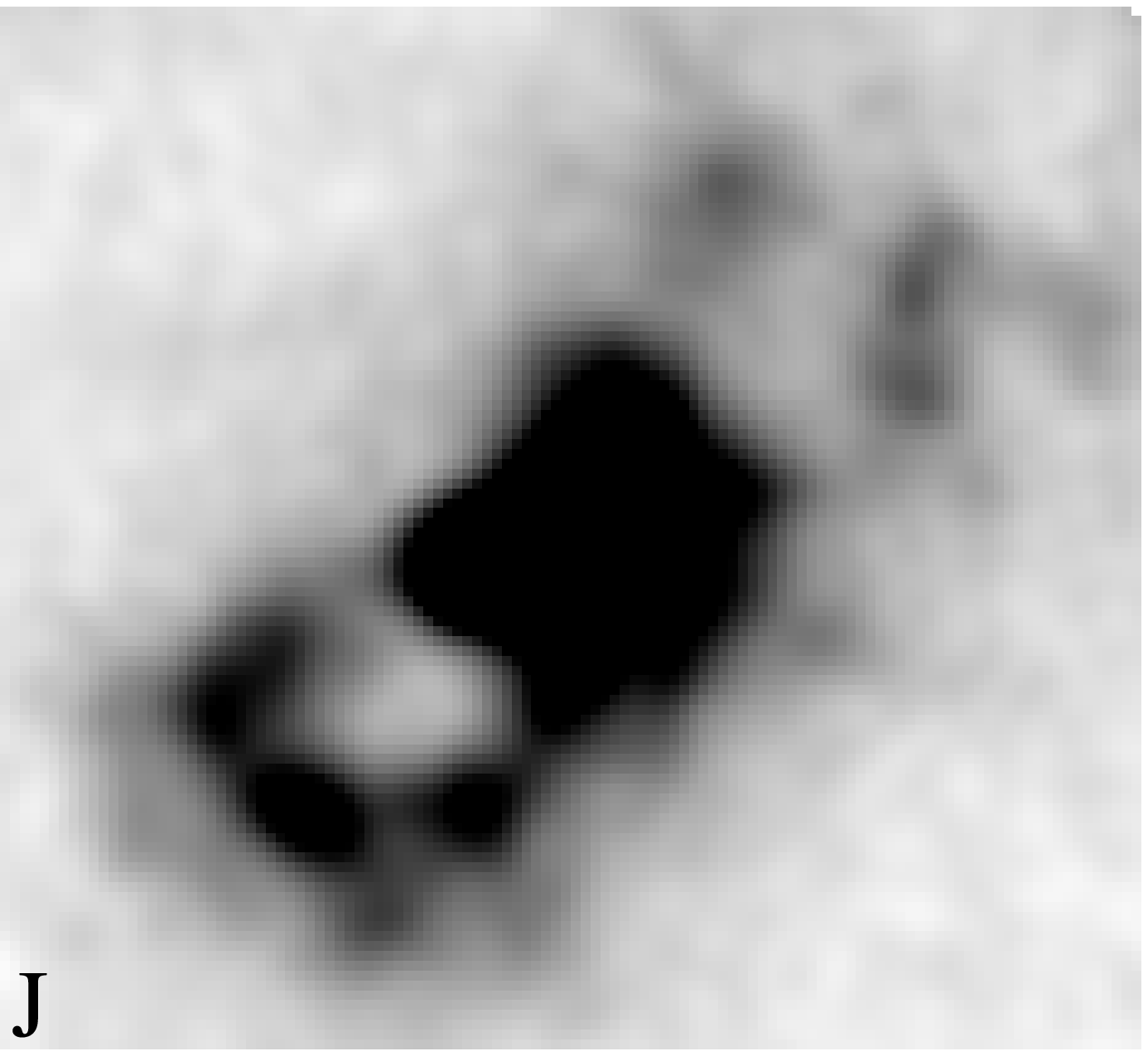}
\includegraphics[height=2.5cm]{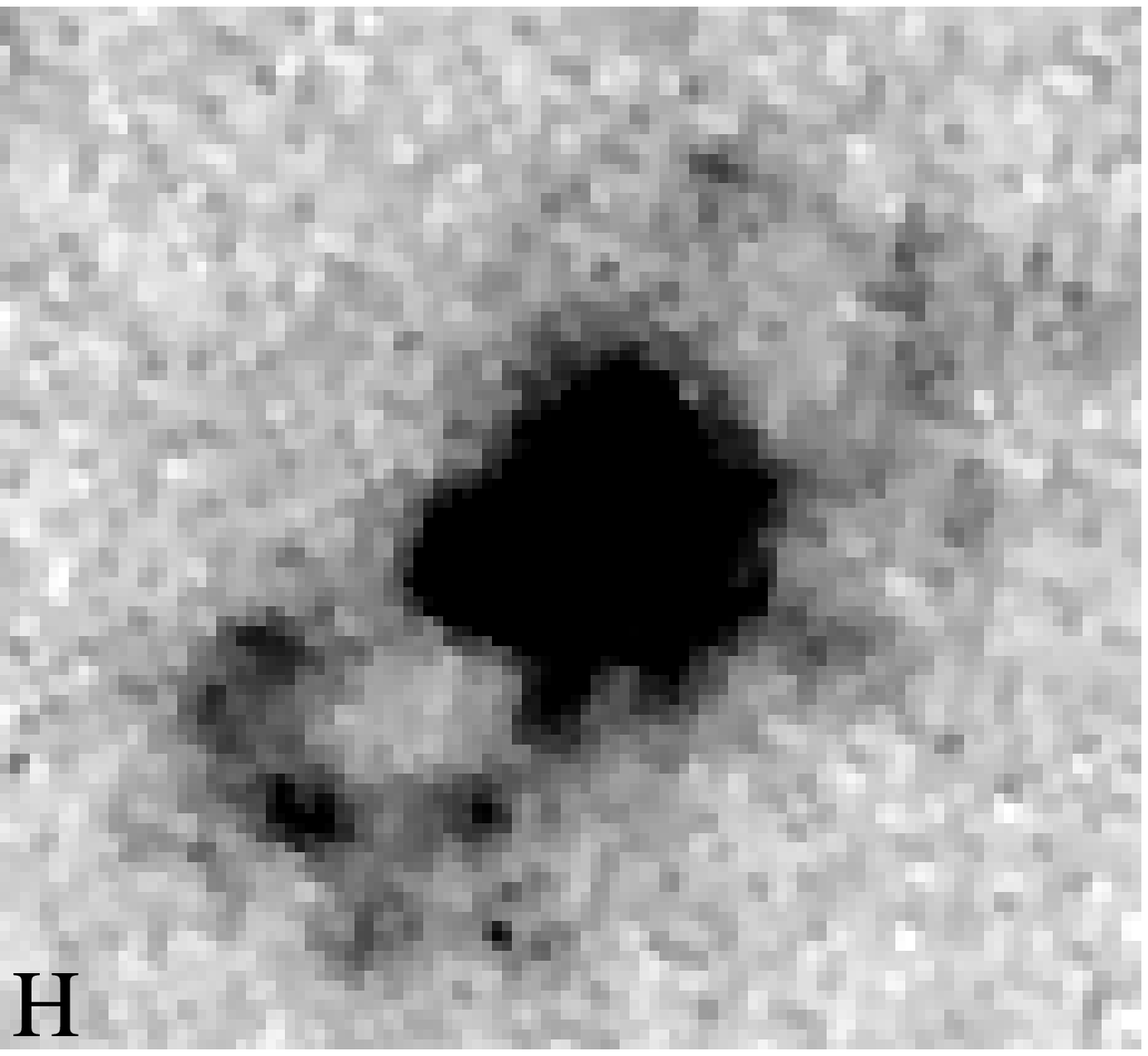}
}
\resizebox{176mm}{!}{
\includegraphics[height=2.5cm]{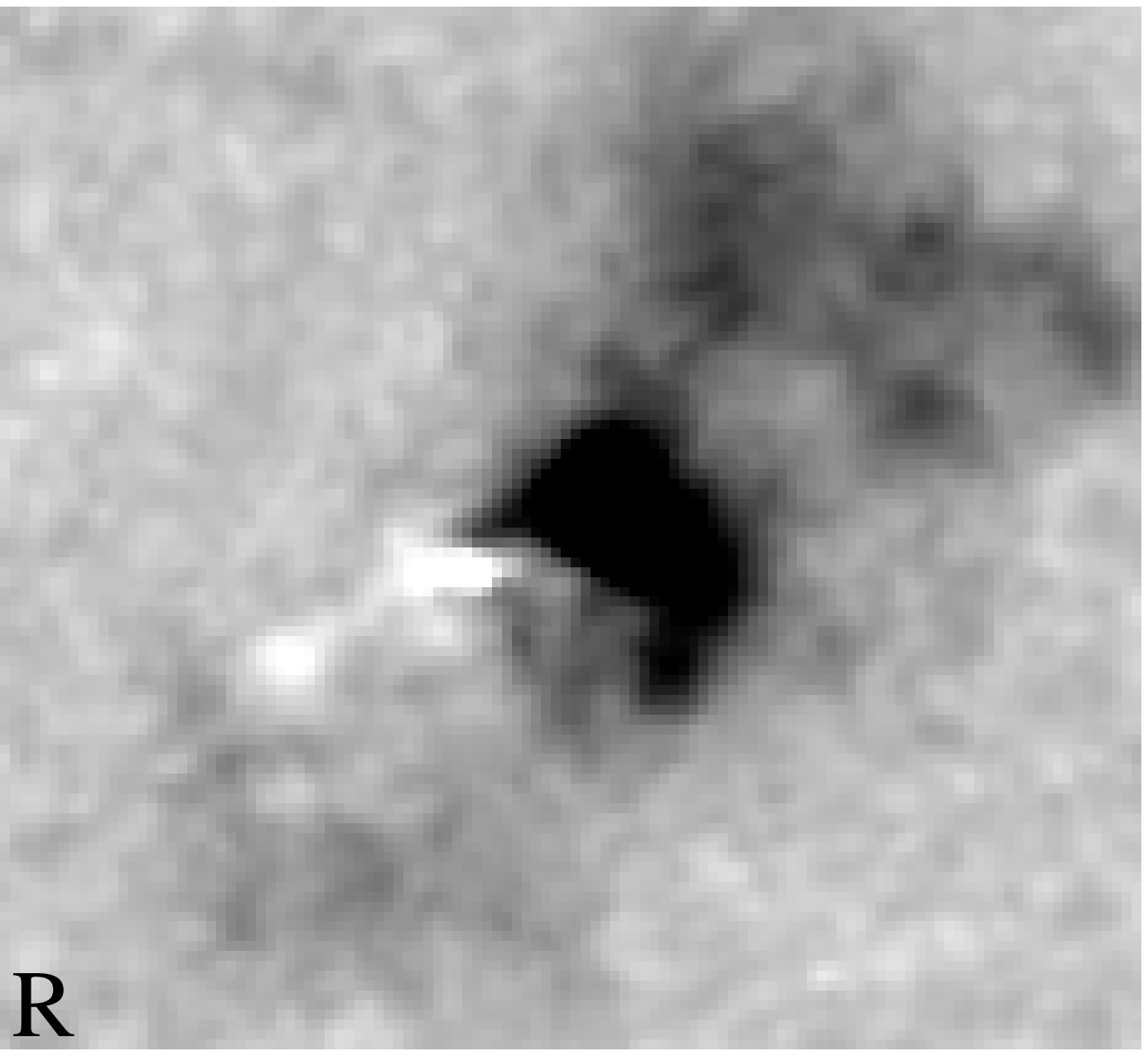}
\includegraphics[height=2.5cm]{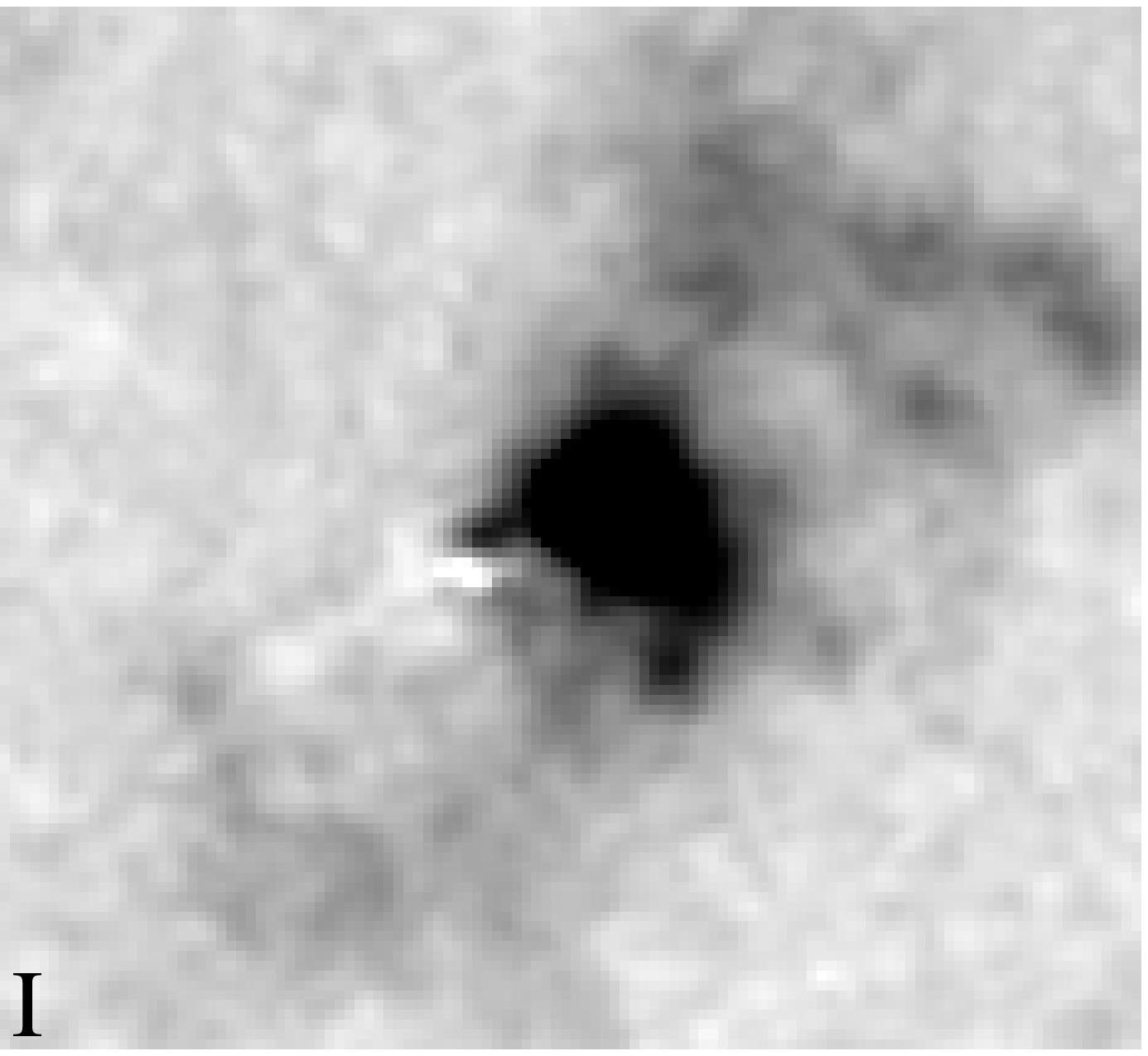}
\includegraphics[height=2.5cm]{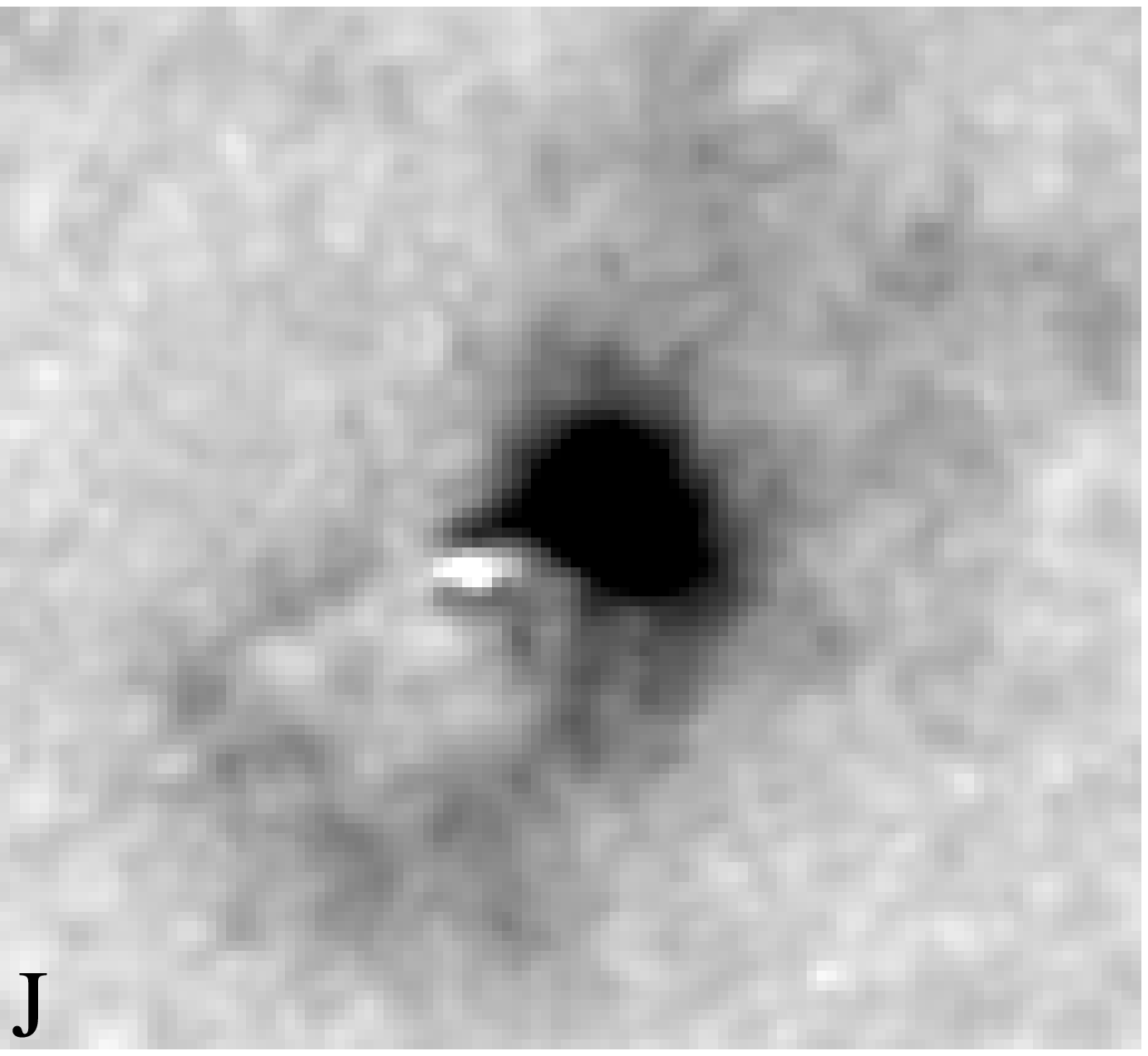}
\includegraphics[height=2.5cm]{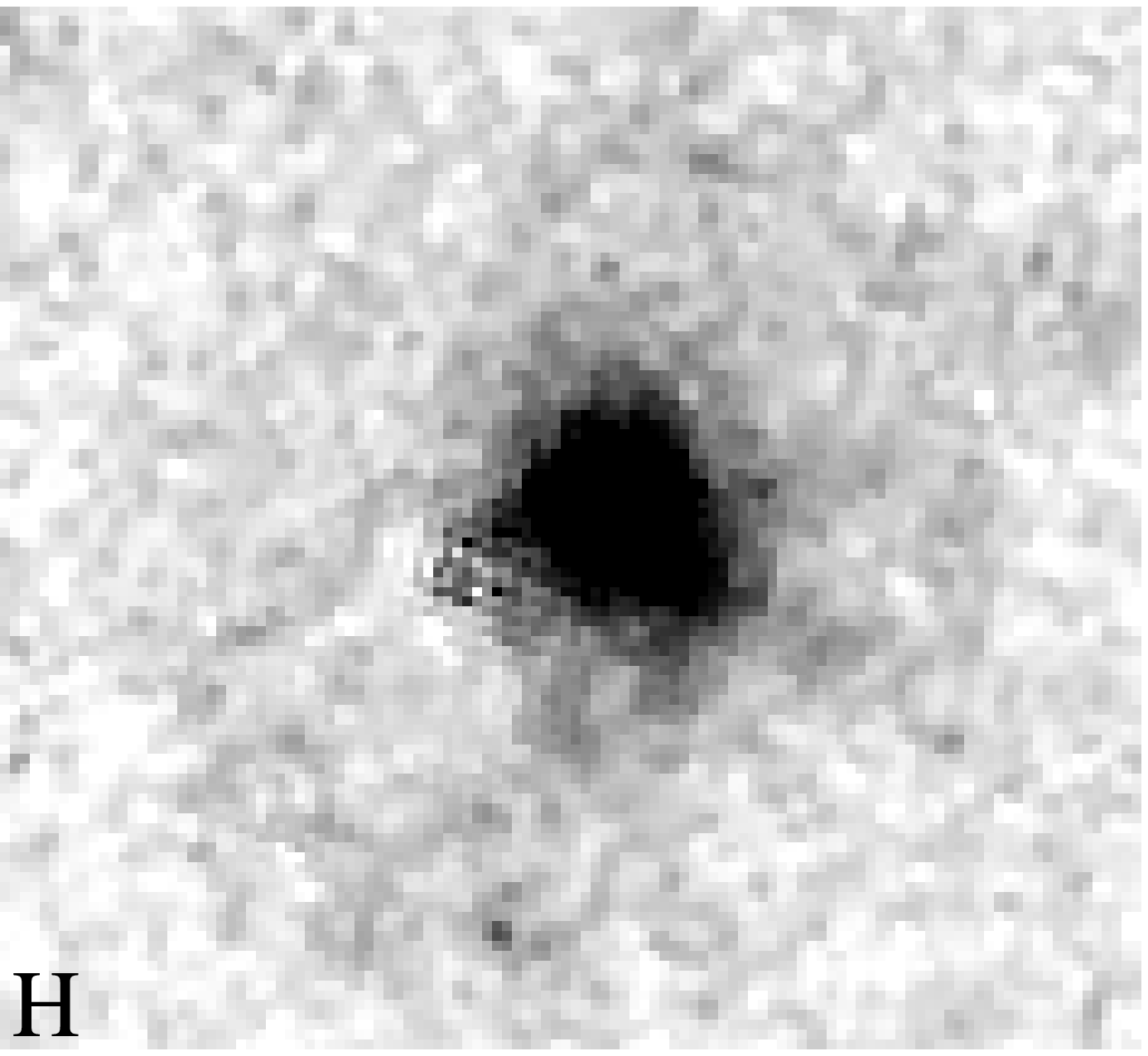}
}
\resizebox{176mm}{!}{
\includegraphics[height=2.5cm]{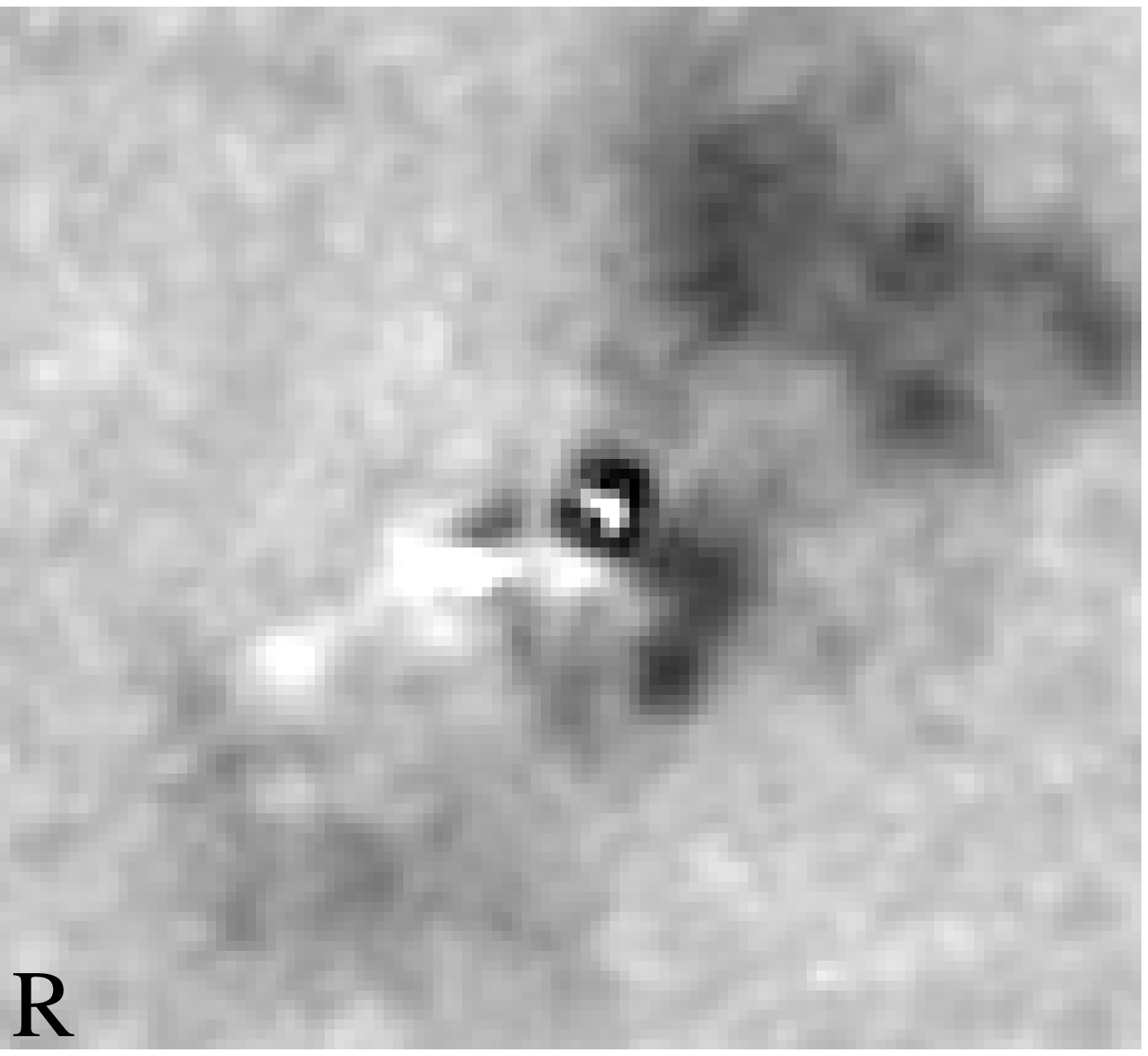}
\includegraphics[height=2.5cm]{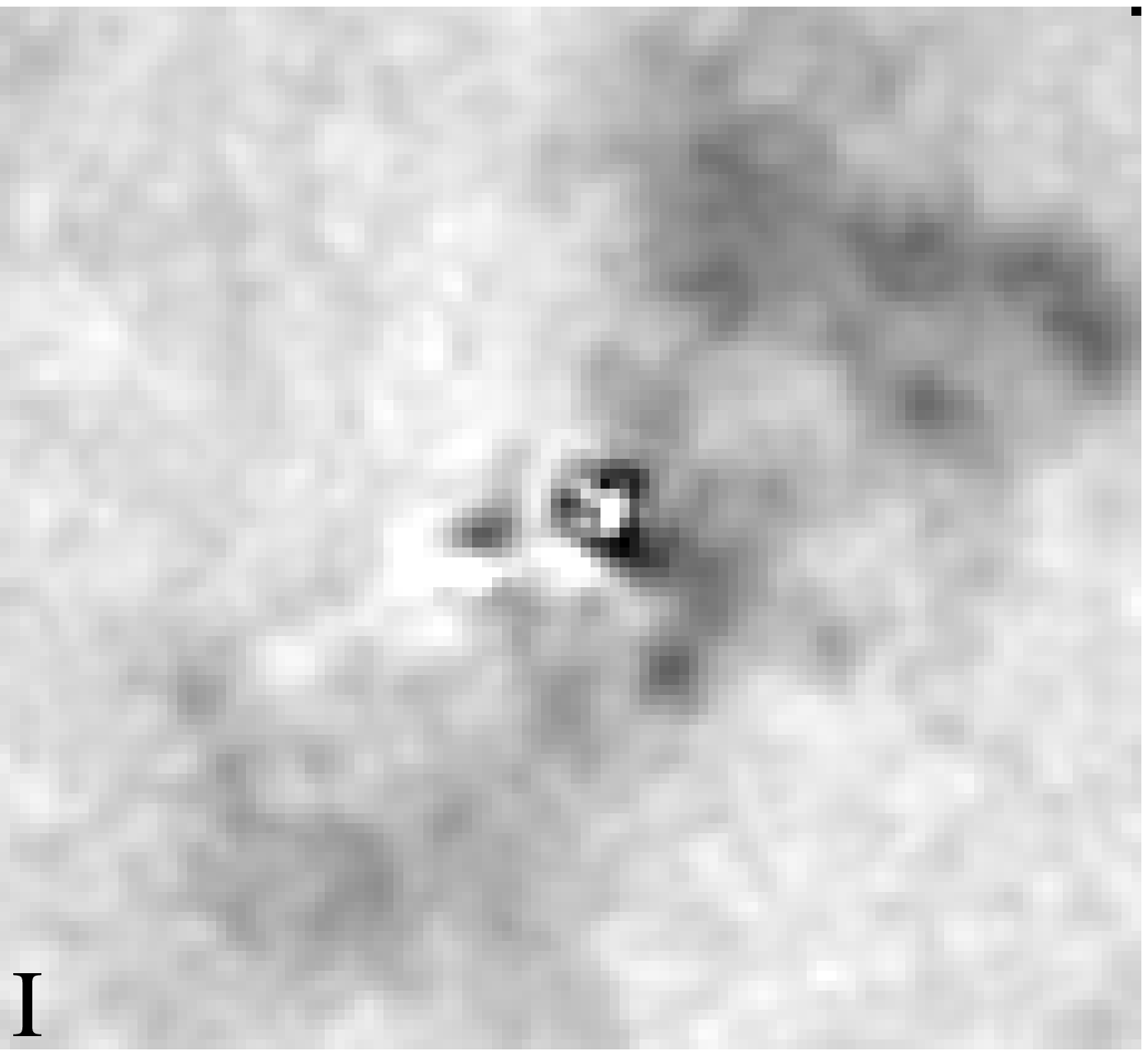}
\includegraphics[height=2.5cm]{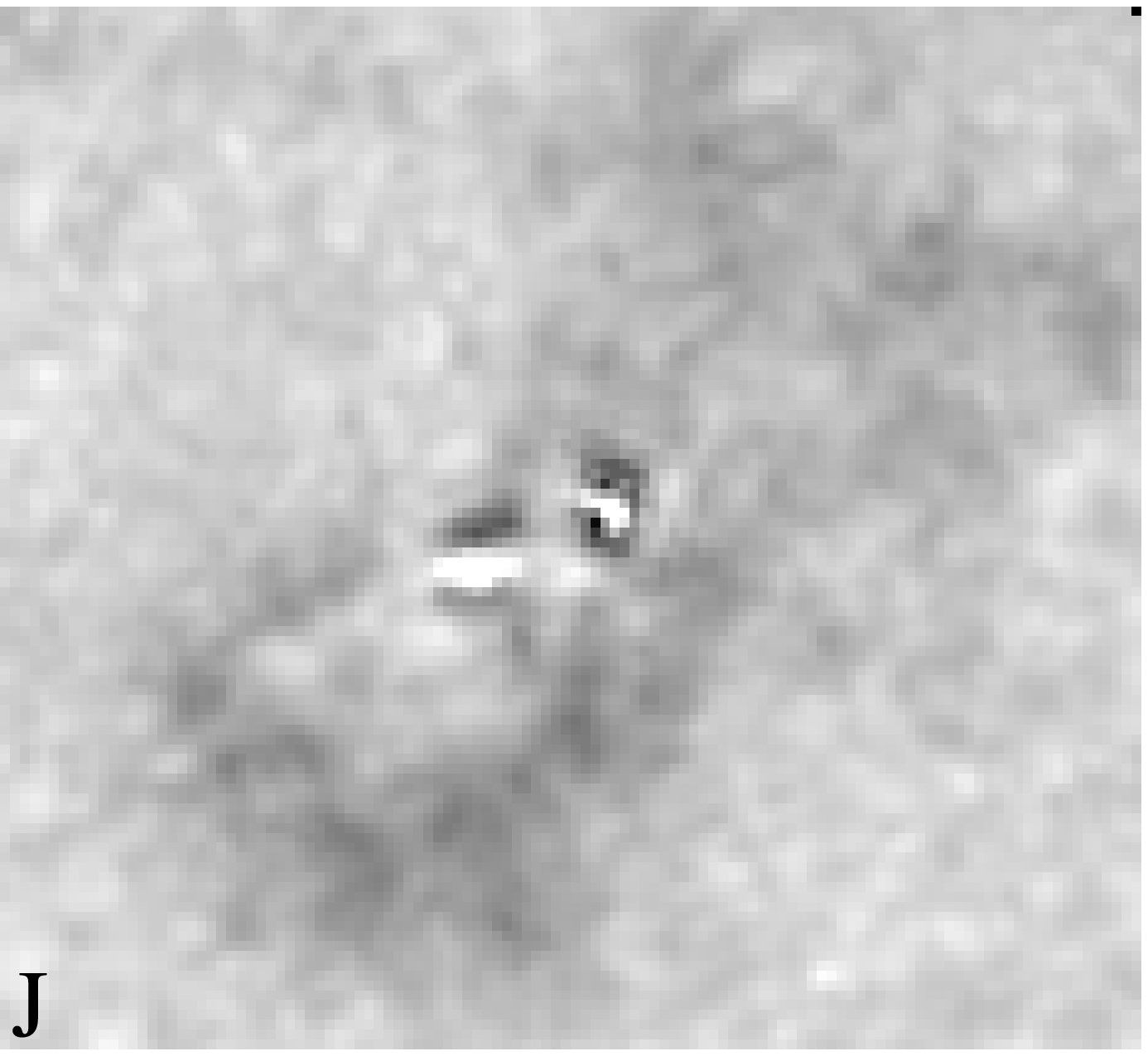}
\includegraphics[height=2.5cm]{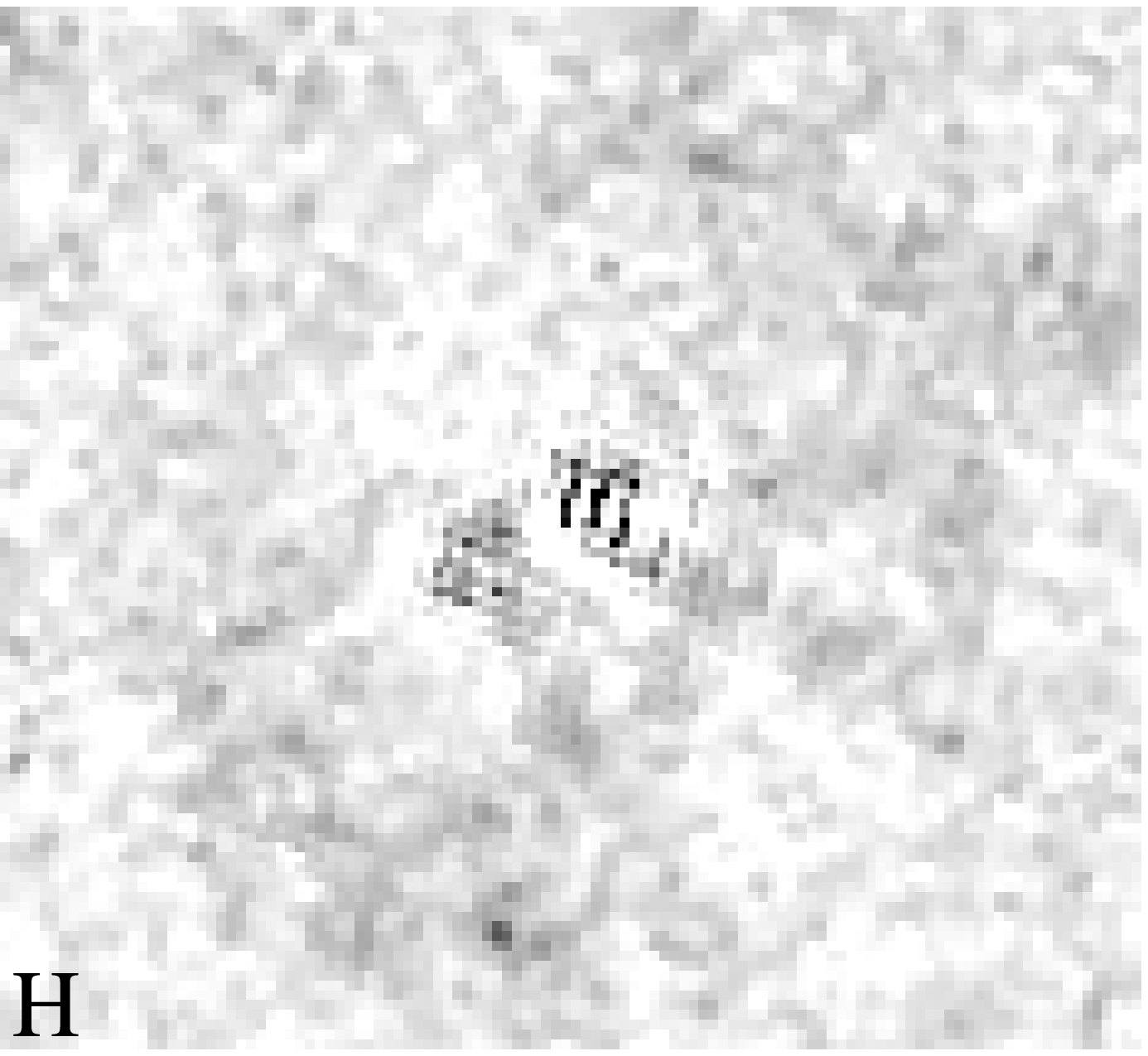}
}
\end{minipage}

\caption{From left to right in all rows: \rf, \iff, \j\ and \h\ {\it HST} images. The top row contain images smoothed to the native resolution of the \h\ WFC3 image. The middle row contain the nebular-subtracted images of \MRC. The bottom row contain the residuals in each image after removing the modelled host galaxy and point-source using {\sc GALFIT}.\label{MRC0406_images}}
\end{figure*}

\subsection{Broad-band continuum images}
\label{neb_sub}
The broad-band images were convolved to match the point spread function (PSF) of the \h\ image (FWHM~$ = 0.15$\,arcsec) using convolution kernels determined with the {\sc iraf} task {\sc psfmatch}.  High frequency noise was removed from the PSF matching function by applying a cosine bell taper in frequency space. The resulting growth curves of stars in the images matched to within 2 per cent at all radii. These smoothed images are presented in the top panels of  Fig.\,\ref{MRC0406_images}.

The nebular emission was then subtracted from the {\it HST} images to reveal continuum-only images of \MRC.  We assume that the nebular emission in the \rf, \iff, \j\ and \h\ images has the same distribution as the [O{\sc iii}] emission. These nebular images were scaled so the light within a $1$\arcsec\ slit centred on the galaxy core and aligned with the radio jets was the percentage of broad-band flux given in  Table \ref{nebular}. The scaled nebular images were then subtracted from the broad-band images. 

The nebular-subtracted images are presented in the middle row of Fig.\,\ref{MRC0406_images}. A dominant central object is clear in all four images.  The bright southeast knot approximately 0.5\arcsec\ from the nucleus is no longer visible in the nebular-subtracted images meaning a large fraction of the flux from this clump was nebular emission. In some images the southeast knot has a negative residual due to over-subtraction of nebular emission, possibly because the line ratios in this knot vary from the global long-slit line-ratios. A large fraction of the extended emission at all wavelengths is due to nebular emission, however, diffuse continuum that extends tens of kpc along the radio jet axis is prominent in the shortest wavelength images.

Before we analyse these images we must highlight the caveats associated with them. The high-resolution images of the [O{\sc iii}] emission show the exact distribution of O$^{++}$ gas only. We derive the ratio of other emission lines relative to [O{\sc iii}] through long-slit spectroscopy, which does not take into account the regional variation of the line ratios. We also make the assumption that the extended emission outside of the slits have the same line-ratios as the emission inside.  Furthermore, a significant fraction of the nebular emission is due to nebular continuum, which has been estimated using a global estimate of the temperature and density of the gas. It is also likely that these properties vary across the nebula, so we expect the nebular continuum contribution to vary spatially as well. Considering these points, it is clear that subtracting the [O{\sc iii}] nebular image from the {\it HST} images is very simplistic, and it is likely that we over- and under-subtract the nebular emission in different regions.

\begin{table*}
\begin{tabular}{|l|c|c|c|c|c|c|c|c|c|}
\hline
&	 \rf\	& \iff\		&\j\ 	&\h\ &	 Mass ($\times10^{10}$\Msun) & SFR (\Msunpyr) & A$_{\rm V}$ & log$_{10}\tau$ & log$_{10}$age  \\  \hline

S\'ersic 		&	23.60$\pm0.17$	&	22.92$\pm0.19$	& 22.34$\pm0.19$	&  21.42$\pm0.19$ &	$10.7^{+9.7}_{-4.8}$	&350$^{+1270}_{-340}$		&$1.5_{-1.3}^{+0.5}$&7.5$^{+2.5}_{-0.3}$		&$8.0_{-0.0}^{+0.9}$\\ 
PS			&	24.78$\pm0.11$	&	24.39$\pm0.11$	&23.83$\pm0.12$	&  23.23$\pm0.12$ &	$1.0^{+2.2}_{-0.3}$	&235$^{+90}_{-220}$		&$1.6_{-0.9}^{+0.2}$&9$^{+1}_{-1.9}$		&$8.0_{-0.0}^{+1.2}$\\ 
Extended 		&	23.0$\pm0.4$		&	22.7$\pm0.5$		&  22.2$\pm0.6$ 	&   $<24.0$	      &	$0.04^{+0.04}_{-0.01}$	&200$^{+1420}_{-110}$		&$0.0_{-0.0}^{+0.4}$&7$^{+3}_{-0}$			&$6.5_{-0.5}^{+0.4}$\\ 
emission	&		&			&  &    \\ \hline
\end{tabular}
\caption{\label{flux_table} The S\'ersic, point-source and extended emission photometry measured by GALFIT by fitting a two component model (a S\'ersic profile and a PSF; see text for details).  All magnitudes are on the AB scale. Object parameters were determined from fitting the photometry with stellar population templates. Uncertainties are quoted at the 68\% level. Photometric uncertainties include a 10\% systematic uncertainty from the flux calibration as well as the uncertainty in the nebular level from Table \ref{nebular}.}
\end{table*}

\section{Analysis}
\label{analysis}
\subsection{Measuring the host galaxy properties with GALFIT}
\label{galfitting}
We measure the morphological properties of the host galaxy using the \h\ band. This filter covers emission beyond rest-frame 4000\AA\ so it is the best tracer of the underlying older stellar population. We fit the nebular-subtracted \h\ image using  \GALFIT\ version 3.0.4 \citep{Peng2010} to derive the morphological properties of the host galaxy. We used a bright unsaturated star in the data as our PSF model. The sky background was measured using {\sc IRAF} and fixed. We fitted a two component model consisting of a PSF with 3 parameters including $x$ position, $y$ position, and total brightness $m_{PSF}$, and a S\'ersic profile \citep{Sersic1968} with 7 free parameters: $x$ position, $y$ position, total brightness $m_{host}$, effective radius $r_e$, S\'ersic index $n$, position angle, and axis ratio.  In total there are 10 free components. All fits were visually checked. A simple single S\'ersic component does not fit the data as the centre of the point-source (PS) component is offset from the centre of the S\'ersic component by $\sim3.2$ pixels, and the residuals from a single S\'ersic fit show clear structure in the core region.

\begin{figure}

\includegraphics[width=1\columnwidth]{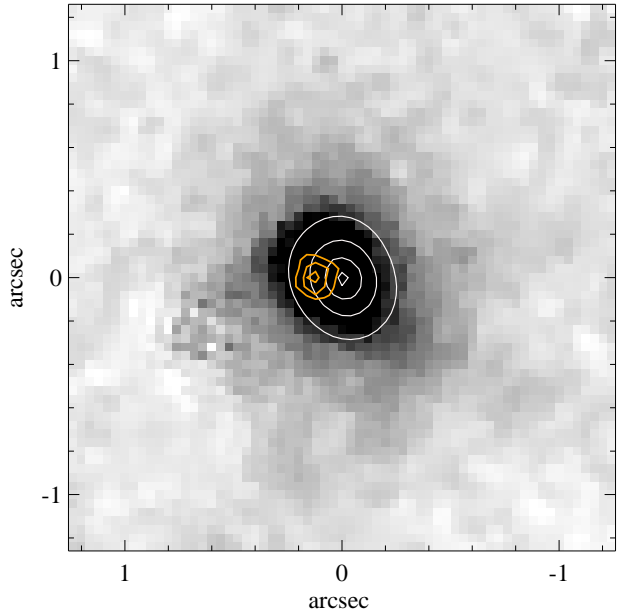}

\caption{A detail of the nebular-subtracted \h\ image of \MRC\ which highlights the location and orientation of the S\'ersic (white contours) and point source components (orange contours). \label{fig:sersic}}

\end{figure}

The input image and residuals of the final iteration of the \GALFIT\ modelling of the nebular-subtracted \h\ image are displayed in the lower two panels of the fourth column of Fig.\,\ref{MRC0406_images}.  Fig.\,\ref{fig:sersic} shows a detail of the nebular-subtracted \h\ image where we have marked the locations and orientations of the best fit point source and S\'ersic components. The S\'ersic component is best fit by an $n=2.2$,  $r_e = 5.7$\,kpc  model, with an axis ratio of 0.75 where the minor axis is $110^\circ$ east of north. However, a range of S\'ersic indexes and effective radii fit the data equally well (resulting in very similar reduced-$\chi^{2}$).  In order to constrain the size of the galaxy we model the S\'ersic component of \MRC\ with both a de Vaucouleur $n=4$ profile as well as a pure disc $n=1$ model. When we fix the S\'ersic index we obtain a minimum $r_e$ of 4.1\,kpc (with $n$ fixed to 1), and a maximum of $r_e=9.7$\,kpc (for $n=4$). We are thus confident that \MRC\ is not a compact source and has effective radius of at least $4.1$\,kpc.  There is only a small difference in the reduced $\chi^2$ of the best-fitting model and these two models, but whilst the residuals of the $n=4$ fit are very similar to the best-fitting model, the $n=1$ model produces a halo of negative residuals surrounding the galaxy core, so \MRC\ is unlikely to be a pure disc galaxy. 

We estimate the flux of the S\'ersic and point-source components in the \rf, \iff\ and \j\ broad-band images using \GALFIT. We fix the morphology and position of the host galaxy and PSF to the best-fit model, but leave the magnitudes of the PSF and S\'ersic components as free parameters. By visually inspecting the fits we ensured the sky background was correctly set so that we did not over-subtract the point-source component in the \rf\ and \iff\ images. 

The photometry of the S\'ersic component, point source, and residual extended emission is given in Table\,\ref{flux_table}.  The uncertainties are derived by performing fits on the nebular-subtracted images created using the upper and lower limits of the nebular fraction from Table \ref{nebular}, then adding in quadrature an additional 10\%  due to uncertainty in the absolute flux calibration. The input images are shown in the middle row of Fig.\,\ref{MRC0406_images}, the final residuals are shown in the bottom row of Fig.\,\ref{MRC0406_images}.  If we force \MRC\ to have a $n=4$ or $n=1$ profile, the total magnitude of the S\'ersic component changes by $\sim0.3$\,mag; the PS component changes by $0.1$\,mag. This magnitude difference does not change the results of the SED fitting in Section \ref{SEDfitting} more than the quoted 1$\sigma$ uncertainties.

\begin{figure*}

\begin{minipage}[h]{176mm}
\centering
\resizebox{176mm}{!}{
\includegraphics[width=\columnwidth]{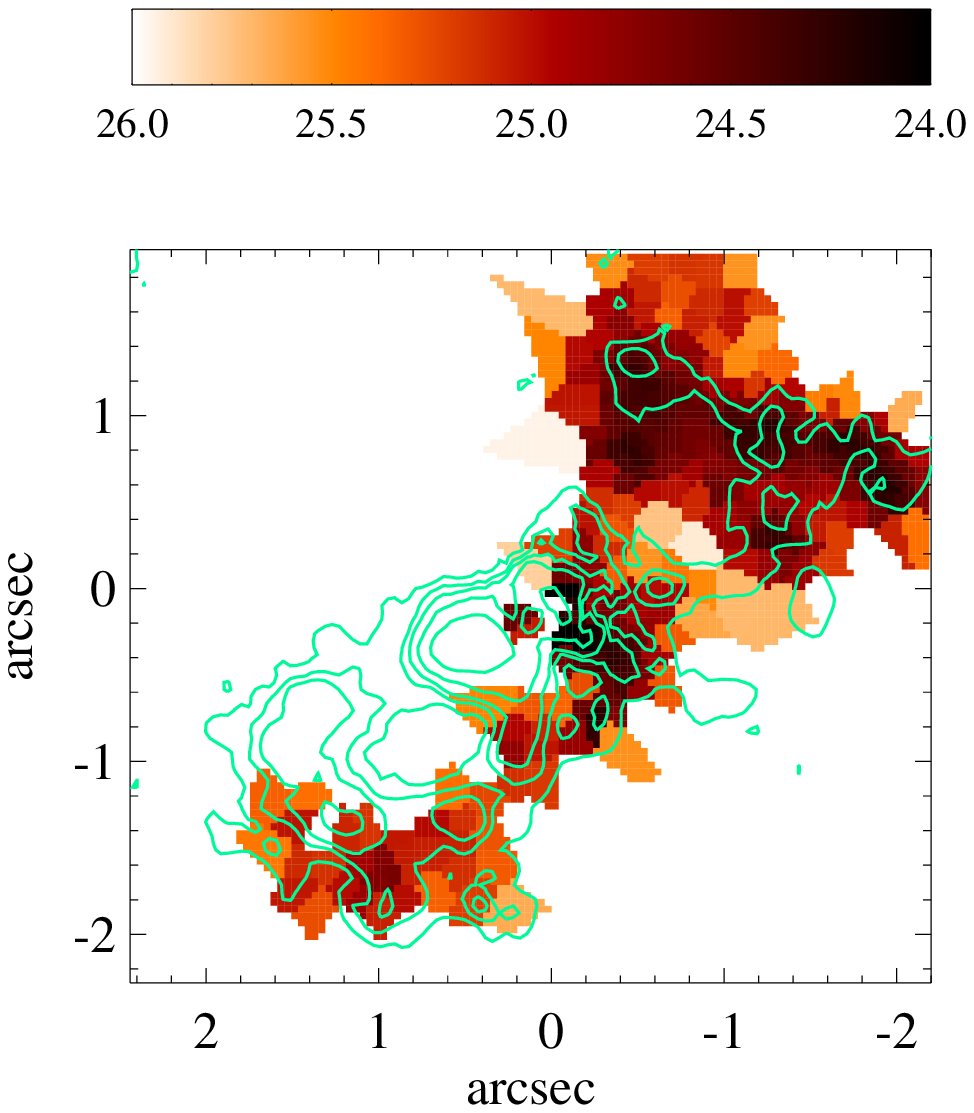}
\includegraphics[width=\columnwidth]{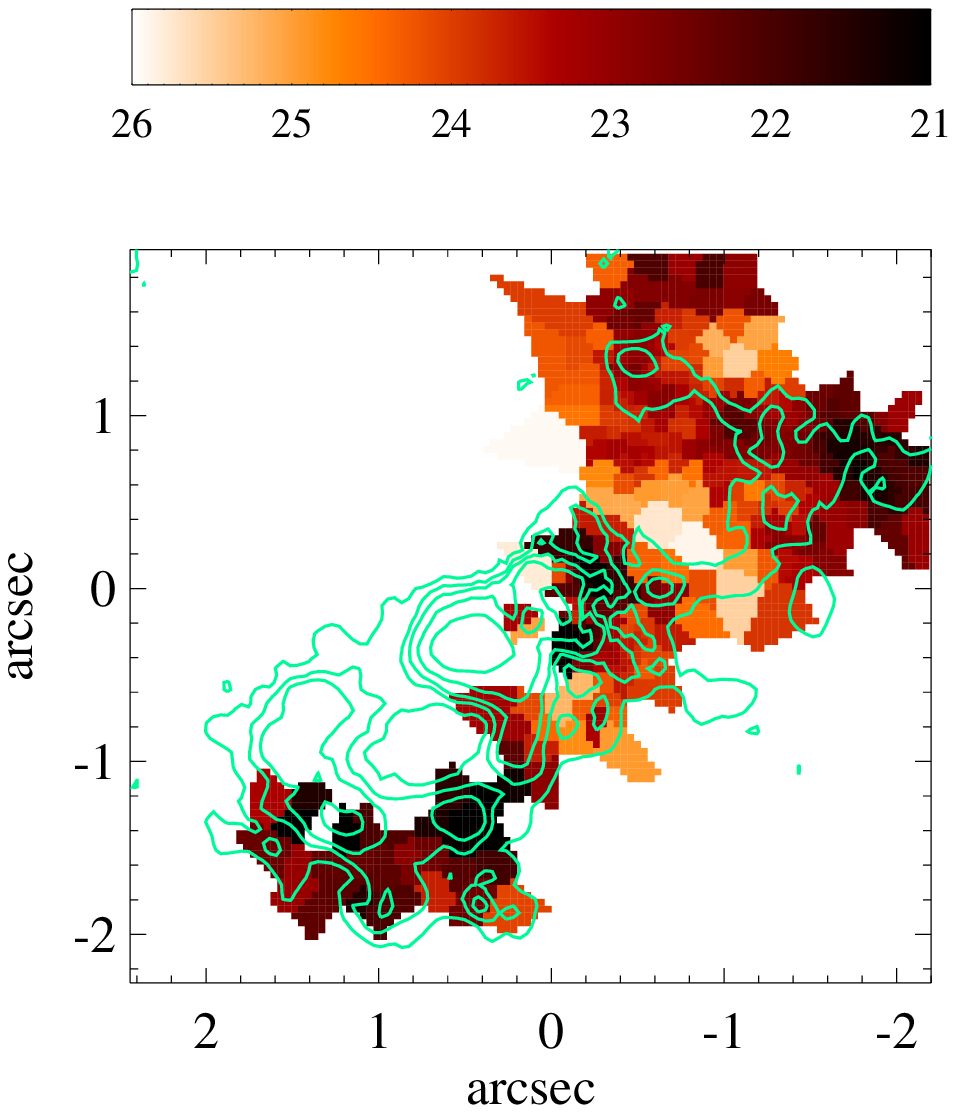}
\includegraphics[width=\columnwidth]{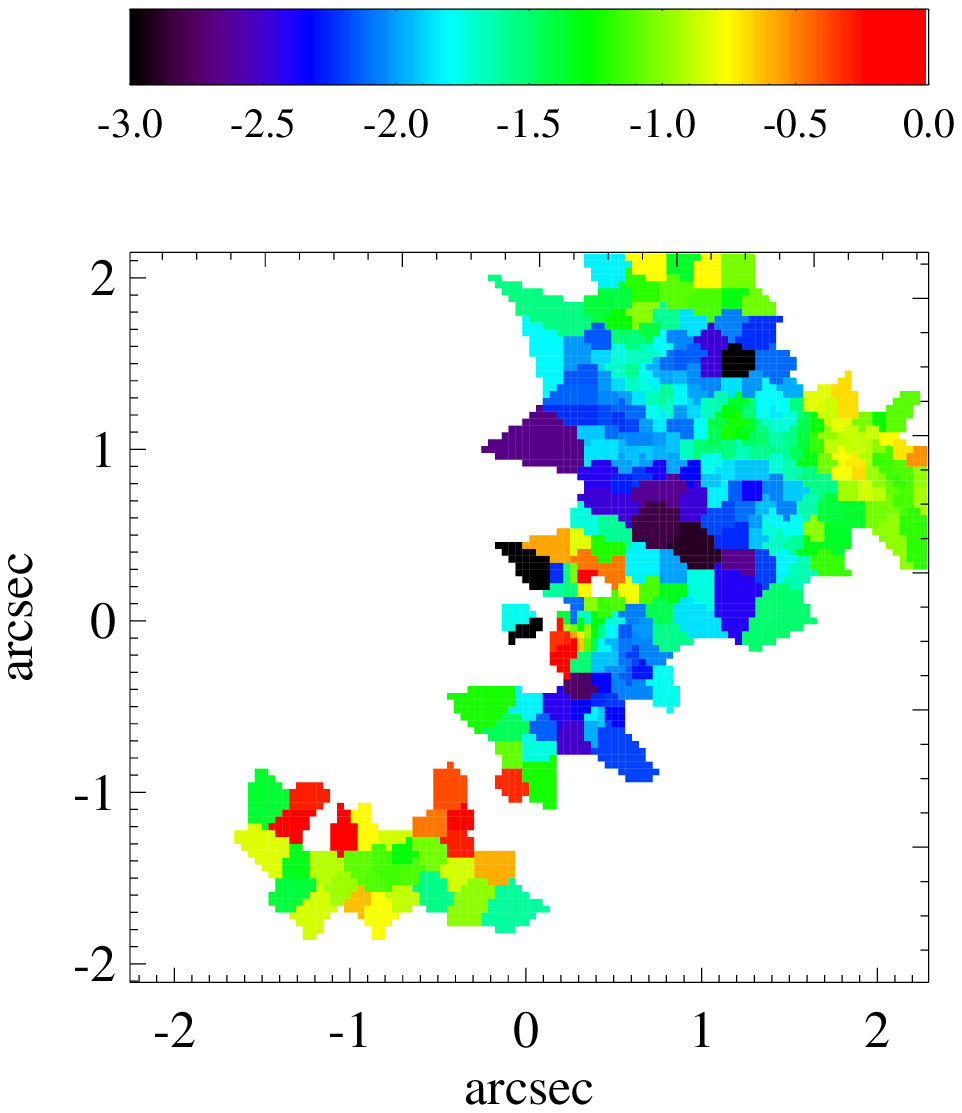}

}
\end{minipage}
\caption{[Left] Comparison of extended \iff\ emission (colour image) and the nebular emission (contours). The colour scale is in mag\,arcsec$^{-2}$ and only bins with \iff\ and \rf $<26\,$mag\,arcsec$^{-2}$ are shown. The contour levels are [1.5, 3, 4.5, 7.5,15]$\times10^{16}$\ergpscmpspsas\ of H$\beta$ flux, or alternatively  [1, 2, 3, 5, 10]\,M$_\odot$yr$^{-1}$kpc$^{-2}$. Both nebular and rest-frame UV emission extend in the same direction out of the central host galaxy, but comparing the spatial correspondence in detail reveals that they are not very similar. [Middle]  After correcting the extended \iff\ emission for dust-extinction there is a much better spatial correspondence with the nebular emission: the southern region is brighter compared to the northern region similar to the nebular distribution [Right] The UV slope, $\beta$, derived from the \rf-\iff\ colour showing a radial colour gradient with the emission becoming redder at large radii. \label{fig:overlay}}

\end{figure*}

The residual images display the extended and clumpy light that cannot be described by a parameterised function. The emission is most prominent in the rest-frame FUV and is approximately biconical. It extends along the radio jet axis in the same direction as the nebular emission and along the minor axis of the host galaxy. However, the distribution of the nebular emission and the extended continuum differs, as shown in the overlay in the left panel of Fig.\,\ref{fig:overlay}. In these figures we have binned the \iff\ data using the Voronoi 2D binning algorithms of \citet{Cappellari2003} to a common signal-to-noise ratio of 20 per bin (see \citealt{Hatch2008} for details on the method and algorithms used).

The right-hand panel of Fig.\,\ref{fig:overlay} maps the UV colour of the extended continuum emission. We plot the slope of the UV continuum, $\beta$, defined such that  $f_{\lambda}=\lambda^{\beta}$. The extended emission is generally blue, and there is a global trend for the extended light to redden with increasing radius: the central and inner region of the northern extension have $\beta=-2.5$ whilst the outer northern and southern regions are redder with $\beta=-0.5$ to $-1$. 

Assuming the red colour of the UV slope is caused by dust reddening, we corrected the \iff\ flux for dust extinction using the \citet{Calzetti2000} extinction law and $E$($B-V$)$=0.22\beta+0.55$ (which is appropriate for a $\sim10^7$\,year old stellar population which has an intrinsic $\beta$ of $-2.5$). The middle panel of Fig.\,\ref{fig:overlay} shows that the extinction-corrected UV emission better resembles the distribution of the nebular emission, with the southern filaments  brighter than the northern filaments, and the bright northern nebular filament now overlapping with a filament of bright UV emission.  It is possible that the nebular and the UV emission trace the same gas and the regions with no detectable UV emission have large extinctions.

\subsection{Spectral energy distribution fitting}
\label{SEDfitting}

It is likely that stellar light contributes significantly to the flux at all wavelengths as the polarisation of the UV emission is less than $5\%$ (J. Vernet, in prep.), which means less than $50\%$ of the UV flux can be scattered AGN light.  We estimate the mass, extinction and star formation rate of the S\'ersic, PS and extended components by fitting the {\it HST} photometry and the {\it Spitzer}-IRAC 3.6$\mu$m and 4.5$\mu$m photometry from \citet{Seymour2007} to stellar population models. 
We use the fitting code FAST \citep{Kriek2009a} with the \citet{Maraston2005} population models and a \citet{Kroupa2001} initial mass function
. We fit delayed exponentially declining (SFR$\sim t e^{-t/\tau}$) star-formation histories with $7<{\rm log}_{10} \tau<10$ in steps of 0.5, $8<{\rm log_{10}\,age}<9.4$\footnote{We allow ${\rm log_{10}\,age}<6$ for the extended emission, however, these young models are highly uncertain and strongly depend on the initial mass function.}  in steps of 0.1 and $0<{\rm A_{V}}<4$ in steps of 0.1\,mag (assuming the \citealt{Calzetti2000} extinction law). The redshift was fixed to 2.44 and the metallicity was fixed to solar abundance.
We do not fit an AGN template to the point source component as \citet{Nesvadba2008} do not detect any broad emission lines from this galaxy, hence it is highly unlikely that this component is reddened AGN light (see Section \ref{ps_discussion} for a fuller discussion).

\begin{figure}
\centering
\includegraphics[width=\columnwidth]{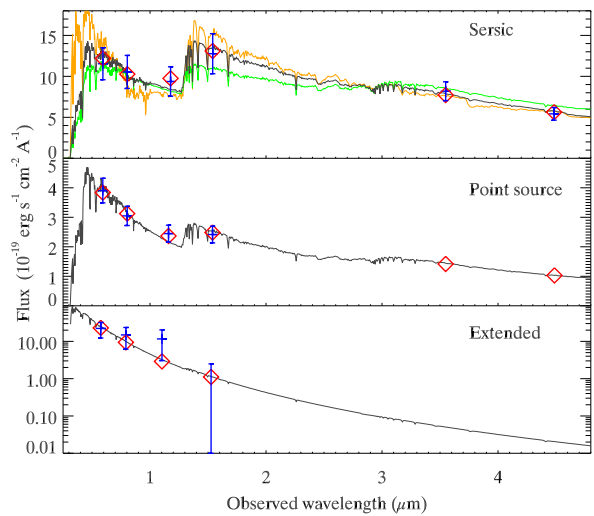}

\caption{SED fit to the S\'ersic component of \MRC. The blue data points show the {\it HST} and {\it Spitzer}-IRAC photometry with 1$\sigma$ uncertainties, the black SED is the best-fit stellar population model to the \rf - IRAC photometry. The red diamonds are the expected photometry through our filters from a galaxy with the plotted SED.  The contribution from the nuclear point-source has been removed from the 3.6$\mu$m and 4.5$\mu$m {\it Spitzer} IRAC photometry.  In the top panel we show 2 additional spectra which cover the range of good fits to the S\'ersic component. The orange spectrum has a star formation rate of $10$\Msunpyr, and the green spectrum has a star formation rate of $1620$\Msunpyr. These spectra show that it is difficult to constrain the star formation rate from these sparsely sampled rest-frame UV to near-infrared data. \label{sed}}
\end{figure}

Approximately 6$\mu$Jy and 7$\mu$Jy of the IRAC 3.6$\mu$m and 4.5$\mu$m flux, respectively, is emitted from the point-source. We assign the remaining 3.6$\mu$m flux of 34.4$\mu$Jy  and 4.5$\mu$m flux of 36.3$\mu$Jy  to the S\'ersic component, and include this IRAC photometry to the stellar population modelling. The properties of the stellar population model that provides the best-fit to the photometry of the S\'ersic, point-source, and extended emission components  are given in Table \ref{flux_table} with 68\% uncertainties. The uncertainties are derived using 100 Monte Carlo simulations in which the photometry is modified according to the uncertainties. In addition, a rest-frame template error function was added to account for uncertainties in the models. See the Appendix in \citet{Kriek2009a} for further details. The templates providing the best-fits to these three components are shown in Fig.\,\ref{sed}.  

Scattered AGN light may contribute to the flux in any of these three components. \citet{Vernet2001} spectropolarimetric study of nine radio galaxies reveals that the SED of the scattered light closely resembles that of quasars. The typical quasar spectrum has a spectral slope of $\beta=-1.56$ between 1350\AA\ and 4230\AA\ \citep{VandenBerk2001}. Therefore scattered AGN light will contribute more to the rest-frame UV data points than the optical and near-infrared data points. This will influence the SED-fitting to select younger models and/or models with lower dust extinction, which will bias the result of the fitting towards lower masses and lower star formation rates (due to the inferred lower extinction).

\section{Discussion}
\label{discussion}
\subsection{Dissecting \MRC}

The light from \MRC\ can be separated into at least 4 components: extended nebular emission, a stellar population distributed in a S\'ersic profile, biconical diffuse continuum emission, and an off-nuclear point-source.

\subsubsection{The host galaxy of \MRC}
The dominating component of \MRC\ in terms of mass is the S\'ersic component. Our SED fitting suggests that this component is a dust obscured, rapidly star-forming, massive ($\sim10^{11}$\,\Msun) galaxy.  The uncertainties of the best-fitting model parameters are large because our sparse photometry is unable to break the degeneracy between age and extinction in the stellar population models. However, we have further evidence that \MRC\ is a dusty galaxy as it is detected by the  {\it Herschel} Space Observatory SPIRE instrument. Drouart et al.\,(in prep) derive the 
starburst-only infrared luminosity  to be $L_{IR}=7.5\pm0.7 \times10^{12} L_\odot$ after deconvolving the AGN and starburst components of the infrared SED. This translates into a star formation rate of $790\pm75$\Msunpyr\ assuming a \citet{Kroupa2001} initial mass function, which is consistent with the rates obtained from our SED fit.

We can robustly place \MRC\ on the $z\sim2$ size-mass relation due to the abundance of recent studies measuring galaxy sizes using \GALFIT\ and WFC3 \h\ data. We convert our measured range of $r_e$ for the optical disc of  \MRC\ to a half light radius ($r_h = r_e(1+q)/2 = 3.5 - 8.2$\,kpc , where $q$ is the axis ratio) and compare it to the size-mass relation of \citet{Newman2012} (see the upper right panel of their Fig.\,3). We find that \MRC\ lies neatly in the region of the mass-size relation that is occupied by star-forming high-redshift galaxies. In fact, in terms of  size and morphology,  \MRC\ is similar to many non-AGN, massive galaxies \citep{Bruce2012}, as well as sub-millimetre galaxies at similar redshifts \citep{Targett2011}.

It is interesting that the radio jet axis ($128^{\circ}$) and galaxy's minor axis ($110^{\circ}$) are relatively similar, as it may suggest that the accretion disc that feeds the black hole is aligned with the host galaxy. Such alignment is not observed in nearby Seyfert galaxies \citep{Kinney2000}, and a much larger sample of high redshift radio galaxies is required to determine whether jet and galaxy axes are typically aligned during this period of coeval rapid black hole and galaxy growth.

\subsubsection{Nature of the point-source component}
\label{ps_discussion}

Two pieces of evidence argue against a quasar origin for the point-source emission: (i) The point-source is offset from the centre of the S\'ersic component by 3.2\,pixels\,$\sim1$\,kpc (see Fig.\,\ref{fig:sersic}); and (ii) \citet{Nesvadba2008} do not detect broad emission lines from the nuclear region. Its photometry is reasonably well fitted with a stellar population of $10^{10}$\Msun\ so it is possible that the point-source is a remnant of a recent merger, or simply a star-forming clump within the disc of the galaxy. Since the point-source is embedded within the S\'ersic component it is difficult to determine whether there is any broadening of the PSF.  Forcing the point-source component to be fit with a S\'ersic profile reveals a very compact ($r_e=0.74$\,kpc), centrally concentrated ($n=9.1$) object, so it is likely that the object is unresolved. This is consistent with it being a star-forming clump, as they typically have radii of a $\sim1$\,kpc at this redshift \citep{Genzel2011}. 

The point-source is not prominent in the [O{\sc iii}] image, however, this does not rule out that the point-source is direct or scattered quasar light. Indeed this system may have a rare off-nuclear AGN, or contain a double AGN. Therefore we cannot make strong conclusions about the origin of this component.

\subsubsection{Origin of the extended emission}

\begin{figure}
\includegraphics[width=0.5\columnwidth]{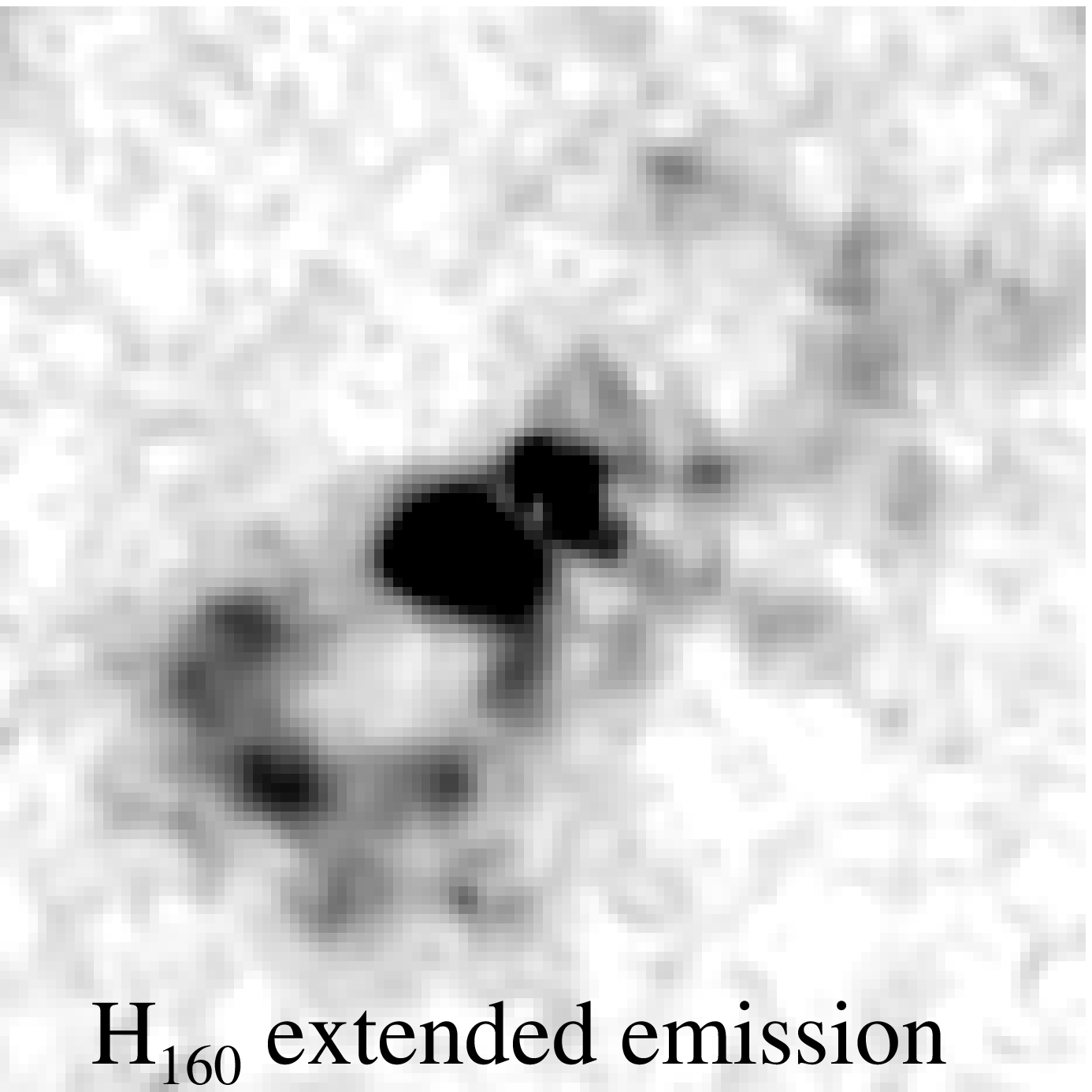}
\includegraphics[width=0.5\columnwidth]{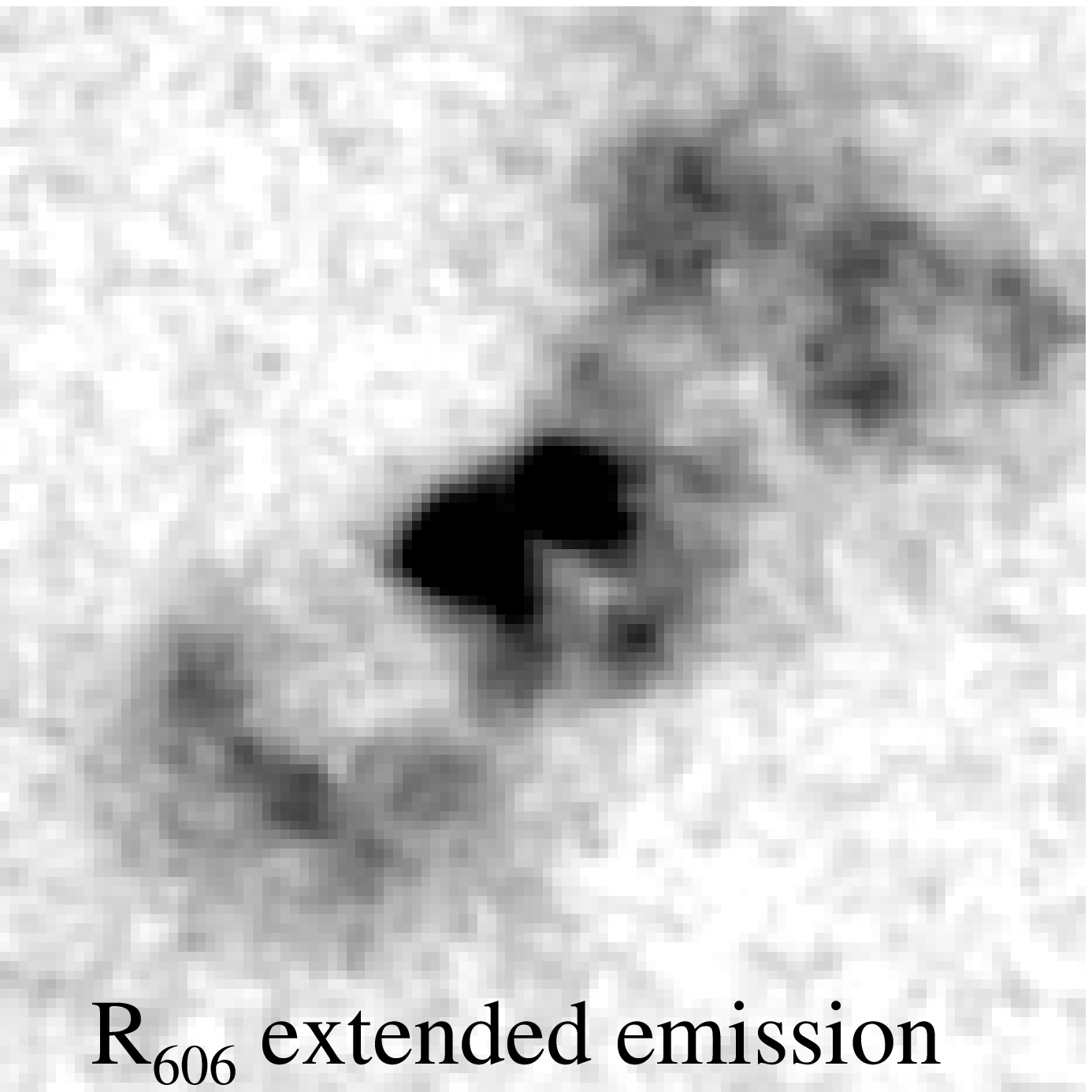}
\caption{A comparison of the extended emission present in the \h\ and \rf\ images. The \rf\ image has been scaled and degraded to match the flux and depth of the \h\ image. The emission in the north is more prominent in the degraded \rf\ image and would have been detected at $>5\sigma$  if present in the \h\ image. Thus the UV-bright extended emission is not nebular emission below the detection limit of the \h\ image. \label{fig:neb_comparison}}
\end{figure}

\label{extended_origin}
Since the rest-frame UV images are deeper than the \h\ images from which we derived the [O{\sc iii}] map we examine the possibility that the UV extended emission is simply low surface brightness nebular emission that is below the detection limits of the \h\ images. We do this by comparing the morphology of the extended emission in the \h\ and \rf\ images after degrading the quality of the \rf\ image to the match that of the \h\ image. If the UV-bright extended emission is due to low surface brightness nebular emission we expect the \h\ and degraded \rf\ images to appear similar. 

Maps of the extended emission were produced by subtracting the best-fitting model of the host galaxy and point source derived in Section \ref{galfitting} from the original  \h\ WFC3 image and the smoothed \rf\ image.  The \rf\ image was then scaled so that the flux of the central 2\arcsec\ region matched that of \h\ extended emission image and noise was added such that the pixel-to-pixel variation also matched. The derived images are compared in Fig.\,\ref{fig:neb_comparison}.

The central and southeastern regions appear similar in both images, however, the emission to the west and northwest is significantly brighter in the \rf\ image than the \h\ image and is detected at $>5\sigma$ significance. If all the extended UV emission was caused by nebular emission then the images would appear more alike and the northern extension would be more prominent in the \h\ image. Thus the UV-bright extended emission seen in the bottom panels of Fig.\,\ref{MRC0406_images} is not nebular emission and must be continuum emission of other origin. Further evidence supporting this conclusion comes from UV spectroscopy.   \citet{Taniguchi2001} shows the emission lines from the southern region are brighter than those in the north, however these regions are of comparable surface brightness in the \rf\ extended emission image; thus at least some of the light from the northern region must be UV continuum. 

The extended continuum emission is unlikely to originate from non-thermal processes such as Synchrotron, Inverse-Compton, and Synchrotron Self Compton.  The strong relation between the electron energy distribution, P, and the flux emitted in the UV means any electron population which has P\,$ < 3$, or equivalently a UV colour $\beta <-1$,  would contribute negligibly to the UV flux (see Section 4.4.2 of \citealt{Hatch2008} for a fuller discussion). Therefore inverse-Compton scattering cannot produce the northern extension. Furthermore, the spectral index of the emission is related to the energy distribution of the electrons doing the scattering, therefore the expected UV colour of any of the above mentioned non-thermal processes will be smooth and not vary on small scales. Our observations show that the UV colour changes drastically on the scales of a kpc (see Fig.\,\ref{fig:overlay}), making such non-thermal origins unlikely.

It is more probable that the extended emission arises from either dust-scattered AGN light, or a diffuse halo of star formation. The polarisation of the UV emission from \MRC\ measured through long-slit spectroscopy is less than $5$ per cent (J. Vernet, in preparation) which means less than half of the total UV flux can result from scattering. From Table \ref{flux_table} we see that $44$ to $68$ per cent of the total UV continuum emitted from \MRC\ comes from the diffuse extended emission, therefore, these polarisation measurements are consistent with some or all of the extended continuum emission resulting from dust-scattered AGN light.

Our SED fitting shows that the photometry of the extended emission is also consistent with it resulting from a young, dust-free stellar population.   The total flux implies a minimum star formation rate of $90$\Msunpyr (and it could be as large as $1620$\Msunpyr). This star formation is spread over a large area so it has a low surface density. 

On the galaxy outskirts the UV emission is redder, with significantly higher $\beta$ values. The UV slope measures the age of the young UV-emitting stellar population, however, it is strongly affected by dust reddening. We are unable to disentangle this degeneracy with our data, however the emission from the outskirts is so red that it cannot be produced by a dust-free old stellar population. A 1\,Gyr old constantly star-forming galaxy would have a maximum $\beta$ of $-2.2$, therefore dust causing at least $A_V=1$\,mag of extinction must be present to redden the light at the outskirts of this system. 

A similar diffuse halo of young stars was also observed around the $z\sim2.2$ radio galaxy MRC\,1138-262 \citep{Hatch2008} and this star formation at large radii can account for the unusually strong Ly$\alpha$ emission around some radio galaxies \citep{VillarMartin2007}. The colour gradient in MRC\,1138-262 is opposite to that observed in \MRC, but is correspondingly so red that the inner part of the halo must contain dust to redden the light.  Such haloes of diffuse star formation are naturally explained by models of star formation driven by AGN feedback \citep[e.g][]{Ishibashi2013}. In such models radiation pressure sweeps up and expels dust from the host galaxy to large radii. This dust compresses gas and forms stars as it moves outward from the galaxy. Biconical winds and outflows are commonly observed in AGN and starburst galaxies, for example, M82. Whilst star-formation within these regions has not previously been reported, radio-triggered star-formation along the jet axis has been observed in the $z=3.8$ radio galaxy 4C41.17 \citep{Dey1997}.

\subsection{Gas and dust beyond the host galaxy}
The nebular and continuum UV light extend up to 25\,kpc from the nucleus of the galaxy, six times the length of the minor axis in the direction of the radio jets. Both extended components contribute to the alignment effect as their distribution is biconical along the jet axis \citep{McCarthy1987,Chambers1987}.  Dust must also exist at similarly large radii as the diffuse blue continuum is either dust-scattered AGN light, or dust-reddened young stars. The dust must have originated in the host galaxy, therefore it is likely that both gas and dust are flowing out of \MRC. Such outflows of dust and metal-rich gas are vitally important for enriching the intergalactic and intracluster medium. 

Nebular gas can be driven out of the galaxy by either expanding starburst bubbles \citep{Humphrey2008}, or overpressurised  cocoons of material surrounding the expanding radio jets \citep{Nesvadba2008}. The dust grains are likely to be destroyed if they interact with the powerful radio jet, so they are likely to have been transported by radiation pressure \citep{Ferrara1990,Nath2009} or supernovae induced winds  \citep{MacLow1999}.

Dust in the intergalactic medium has long been theorised, but only recently detected. \citet{Menard2010} inferred that half of the cosmic dust lies within galaxies, whilst the other half has been expelled and lies in the intergalactic medium. The extended light in \MRC\ allows us a rare opportunity to observe this dust in the process of leaving the galaxy. Since the host galaxy of \MRC\ does not appear different to other massive star-forming galaxies, it is possible that many highly star-forming galaxies at a similar redshift also expel similar amounts of dust in bipolar outflows. This dust will remain invisible unless it is illuminated by a sufficiently bright AGN or feedback driven star formation. If such dusty outflows are common we may expect to see UV-bright emission along the minor axis of stacked images of starbursting AGN.

\subsection{Has \MRC\ undergone a recent merger?}

Gas rich mergers are commonly found in high-redshift radio galaxies suggesting that an interaction with a gas rich companion may drive the star formation and nuclear activity \citep{Ivison2012}. \MRC\ was chosen for this study as it has one of the most complex galaxy morphologies, consisting of multiple clumps and filaments that extend tens of kpc from the galaxy core. By separating the broad-band light into various components, we aimed to determine whether the clumpy structure of this radio galaxy is a signature that it has undergone a recent merger. 

Several previous studies have already considered this issue. \citet{Rush1997} concluded that the extended continuum and the emission-line kinematics of \MRC\ suggested that this galaxy had recently undergone a merger. More recent work interpreted the same emission-line kinematics as radio-jet AGN feedback \citep{Nesvadba2008}, or a starburst driven wind \citep{Taniguchi2001,Humphrey2009}. \citet{Nesvadba2008} provided the strongest evidence against a recent major merger as they separated the continuum emission from the nebular emission using a near-infrared integral field spectrograph to reveal only a single continuum source. However, the low spatial resolution of their data could not separate sources that were less than 5\,kpc apart.

Using our new {\it HST} images we show that there are two continuum components in \MRC\ that lie 1.1\,kpc apart in projection: a smooth S\'ersic profile plus a point-source.  It is possible that the point-source component is the remanent from a recent gas-rich merger which may have triggered the current activity. Two or more distinct components have been observed in other radio galaxies, for example \citet{Ivison2012} found multiple submillimetre components indicative of gas-rich mergers and \citet{Rocca-Volmerange2013} required two stellar components to adequately fit the UV to submillimetre SEDs of radio galaxies, so it is plausible that this point-source is a separate galaxy remnant.

However, star-forming clumps of the same size as the point-source are common within the discs of star-forming galaxies (although its mass is comparatively high; \citealt{Wuyts2012}).  So the presence of this component does not imply that a merger has occurred. In addition, simulations of galaxy mergers suggest that smooth stellar structures only re-form more than $100$\,Myr after the final coalescence, so the smooth S\'ersic component of \MRC\ argues against a very recent major-merger  \citep{Robertson2006}. 

We conclude that the most disruption that \MRC\ could have recently undergone is a minor merger since the point source has a 1:10 mass ratio with the host galaxy, and the stellar disc is relatively smooth. Therefore recent major-mergers are not required to trigger the strong AGN or sub-millimetre activity, and the stellar component of the host galaxy can grow rapidly without disrupting its original shape.

\subsection{Future evolution of \MRC}
We now speculate on what type of galaxy \MRC\ will evolve into. The extended, clumpy structure of \MRC\ makes it appear peculiar and irregular, but both the extended nebular and continuum emission are produced by short-lived processes.  \citet{Humphrey2009} argue that the nebular gas will disperse into the intergalactic medium in $\sim100$\,Myrs based on the emission-line kinematics. In addition, when the AGN stops accreting and decreases its luminosity, the surrounding gas will no longer be ionized and will rapidly become invisible. If the extended continuum is produced by dust-scattered AGN light, then this too will disappear when the AGN halts. If it is a halo of forming stars, then it may continue to be visible for the duration of the starburst phase, which is also expected to be relatively short-lived ($<$100\,Myr). Thus the alignment effect for \MRC\ is a short-lived phenomenon, and once it is over,  \MRC\ may have a similar appearance to typical massive, star-forming galaxies. 

Whilst \MRC\ is a rapidly star-forming galaxy at the observed epoch, its AGN activity is strongly affecting the host galaxy. \citet{Nesvadba2008} suggest that ionized gas is being entrained out of the galaxy by the radio jets at a rate of $100-1000$\Msunpyr\ over the $\sim10$\,Myr lifetime of the radio source. In addition to this, the large sub-millimetre flux and our SED fits suggest that gas is being locked into stars at similar high rates, so the gas reservoir of \MRC\ is depleting rapidly. We therefore expect the star formation in \MRC\ to slow down and perhaps reduce to such low levels that it may be classified as a passive galaxy.

We speculate that the feedback-driven star formation we observed at large radii may alter the shape of the galaxy. \citet{Ishibashi2013} suggest that the stars formed in the diffuse halo would remain on radial orbits. This would act to increase the size of the host galaxy. Our observations of \MRC\ show that the AGN-driven star formation occurs in biconical shells that are orientated along the minor axis of the host galaxy. Thus these stars are likely to puff-up the minor axis of the galaxy preferentially and can not only increase the galaxy's size, but also alter its morphology such that the stellar profile grows into a more spheroidal shape.

\section{Summary}
We have dissected the powerful $z=2.4$ radio galaxy \MRC\ that is undergoing AGN feedback. We have separated the emission into the host galaxy, a central point-source, nebular emission, and extended continuum emission. We present high resolution [O{\sc iii}] and H$\beta$ images showing that most of the clumps and extended filamentary features in the broad-band images are due to nebular emission that has been ionized by the central AGN.

The host galaxy is a dust enshrouded, massive galaxy (M$_\star\sim10^{11}$\Msun), with a high submillimetre star-formation rate ($790\pm75$\Msunpyr). Its morphology is well described by a  S\'ersic component, with a half-light radius in the range  $3.5-8.2$\kpc, we therefore conclude that the progenitor of this radio galaxy was a massive star-forming galaxy. The stellar disc is intact which argues against a recent major merger, however, the central point-source may be the remnant of a gas-rich minor merger.

Our images also reveal intriguing biconical diffuse emission, extending up to $25$\,kpc along the radio jet axis, which is likely to arise from dust-scattered AGN light or dusty star formation. Regardless of the illumination source, this emission reveals that gas and dust has been expelled beyond several optical half-light radii by the AGN and starburst activity, and pollutes the surrounding intergalactic medium.

\section{Acknowledgments}
We sincerely thank the referee for providing useful and constructive comments which improved this paper. We also thank Nick Seymour for providing the NICMOS image of \MRC, Boris H\"aussler for help with GALFIT, and Dan Smith for useful discussions. NAH is supported by an STFC Rutherford Fellowship and the University of Nottingham Anne McLaren Fellowship. JDK thanks the DFG for support via German-Israeli Project Cooperation grant
STE1869/1-1.GE625/15-1.
This work is based on observations made with the NASA/ESA Hubble Space Telescope, obtained at the Space Telescope Science Institute, which is operated by the Association of Universities for Research in Astronomy, Inc., under NASA contract NAS 5-26555.
\bibliographystyle{mn2e}\bibliography{HzRG-bib,mn-jour}
\label{lastpage}
\clearpage
\end{document}